\documentclass[sigconf]{acmart}

\usepackage{colortbl}
\usepackage{multirow}
\usepackage{subfig}
\usepackage{graphicx}
\usepackage{makecell}
\usepackage{appendix}
\usepackage{booktabs}
\usepackage{balance}
\usepackage{longtable}
\usepackage{cancel}
\usepackage{xcolor}
\usepackage{fontawesome}
\usepackage[utf8]{inputenc}

\DeclareUnicodeCharacter{2082}{\ensuremath{{}_2}}

\AtBeginDocument{%
  \providecommand\BibTeX{{%
    \normalfont B\kern-0.5em{\scshape i\kern-0.25em b}\kern-0.8em\TeX}}}

\copyrightyear{2024}
\acmYear{2024}
\setcopyright{acmlicensed}\acmConference[CHI '24]{Proceedings of the CHI Conference on Human Factors in Computing Systems}{May 11--16, 2024}{Honolulu, HI, USA}
\acmBooktitle{Proceedings of the CHI Conference on Human Factors in Computing Systems (CHI '24), May 11--16, 2024, Honolulu, HI, USA}
\acmDOI{10.1145/3613904.3642813}
\acmISBN{979-8-4007-0330-0/24/05}

\newcommand{\add}[1]{\textcolor{black}{#1}}

\makeatletter
\let\orgdescriptionlabel\descriptionlabel
\renewcommand*{\descriptionlabel}[1]{%
  \let\orglabel\label
  \let\label\@gobble
  \phantomsection
  \edef\@currentlabel{#1\unskip}%
  \let\label\orglabel
  \orgdescriptionlabel{#1}%
}
 \makeatother

\begin{document}

\title{Do You See What I See? A Qualitative Study Eliciting High-Level Visualization Comprehension}

\author{Ghulam Jilani Quadri}
\email{jiquad@cs.unc.edu}
\orcid{0000-0002-8054-5048}
\affiliation{%
  \institution{University of North Carolina at Chapel Hill}
  \city{Chapel Hill, NC}
  \country{USA}}

\author{Arran Zeyu Wang}
\email{zeyuwang@cs.unc.edu}
\orcid{0000-0002-7491-7570}
\affiliation{%
  \institution{University of North Carolina at Chapel Hill}
  \city{Chapel Hill, NC}
  \country{USA}}

\author{Zhehao Wang}
\email{zhehaow@email.unc.edu}
\orcid{0009-0009-8490-5491}
\affiliation{%
  \institution{University of North Carolina at Chapel Hill}
  \city{Chapel Hill, NC}
  \country{USA}}

\author{Jennifer Adorno}	
\email{jadorno4@usf.edu}
\orcid{0000-0002-6511-694X}
\affiliation{%
  \institution{University of South Florida}
  \city{Tampa, FL}
  \country{USA}}

\author{Paul Rosen}
\email{paul.rosen@utah.edu}
\orcid{0000-0002-0873-9518}
\affiliation{%
 \institution{University of Utah}
   \city{Salt Lake City, UT}
  \country{USA}}

\author{Danielle Albers Szafir}
\email{danielle.szafir@cs.unc.edu}
\orcid{0000-0003-3634-8597}
\affiliation{%
  \institution{University of North Carolina at Chapel Hill}
  \city{Chapel Hill, NC}
  \country{USA}}

\renewcommand{\shortauthors}{Quadri, et al.}

\begin{abstract}

Designers often create visualizations to achieve specific high-level analytical or communication goals. These goals require people to naturally extract complex, contextualized, and interconnected patterns in data. While limited prior work has studied general high-level interpretation, prevailing perceptual studies of visualization effectiveness primarily focus on isolated, predefined, low-level tasks, such as estimating statistical quantities. This study more holistically explores visualization interpretation to examine the alignment between designers' communicative goals and what their audience sees in a visualization, which we refer to as their \textit{comprehension}. We found that statistics people effectively estimate from visualizations in classical graphical perception studies may differ from the patterns people intuitively comprehend in a visualization. We conducted a qualitative study on three types of visualizations---line graphs, bar graphs, and scatterplots---to investigate the high-level patterns people naturally draw from a visualization. Participants described a series of graphs using natural language and think-aloud protocols. We found that comprehension varies with a range of factors, including graph complexity and data distribution. Specifically, 1) a visualization's stated objective often does not align with people's comprehension, 2) results from traditional experiments may not predict the knowledge people build with a graph, and 3) chart type alone is insufficient to predict the information people extract from a graph. Our study confirms the importance of defining visualization effectiveness from multiple perspectives to assess and inform visualization practices.

\end{abstract}



\begin{teaserfigure}
\centering
\includegraphics[width=1\columnwidth]{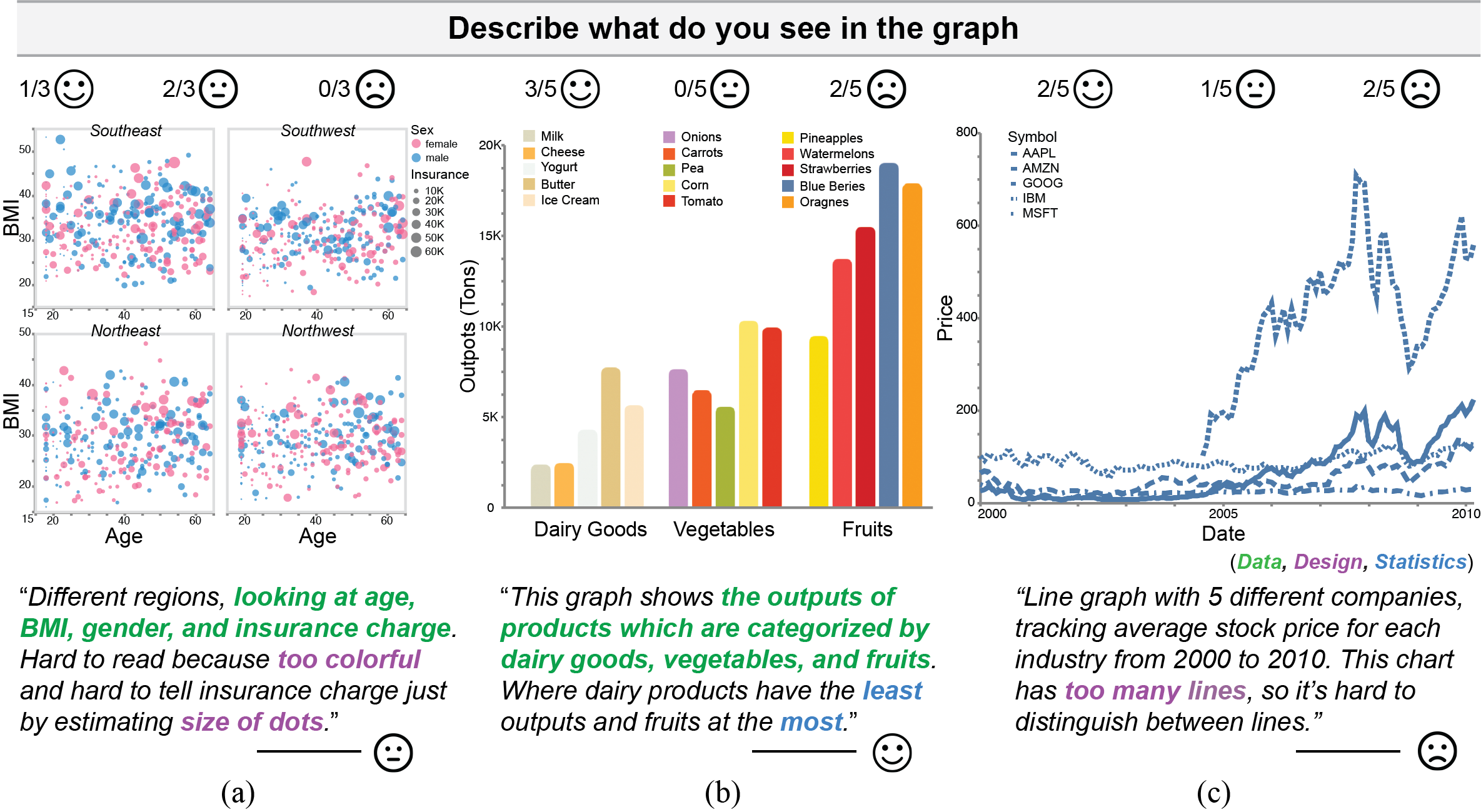}
\caption{Sample responses participants provided when asked to describe the contents of three graphs:
(a) multi-class juxtaposed scatterplot,
(b) multi-class juxtaposed bar graph,
and (c)
multi-class line graph.
Happy faces 
\begingroup\normalfont
 \includegraphics[height=7pt]{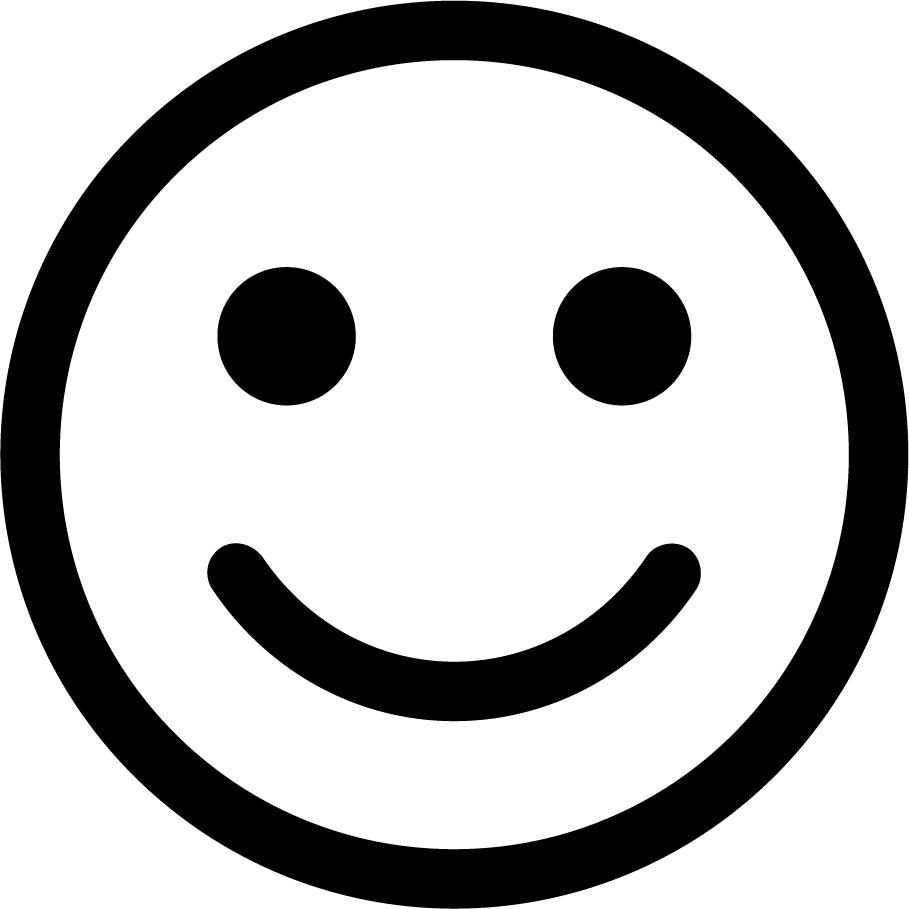}%
\endgroup{} 
indicate the number of responses that match the designers' intended communication goals, neutral faces 
\begingroup\normalfont
 \includegraphics[height=7pt]{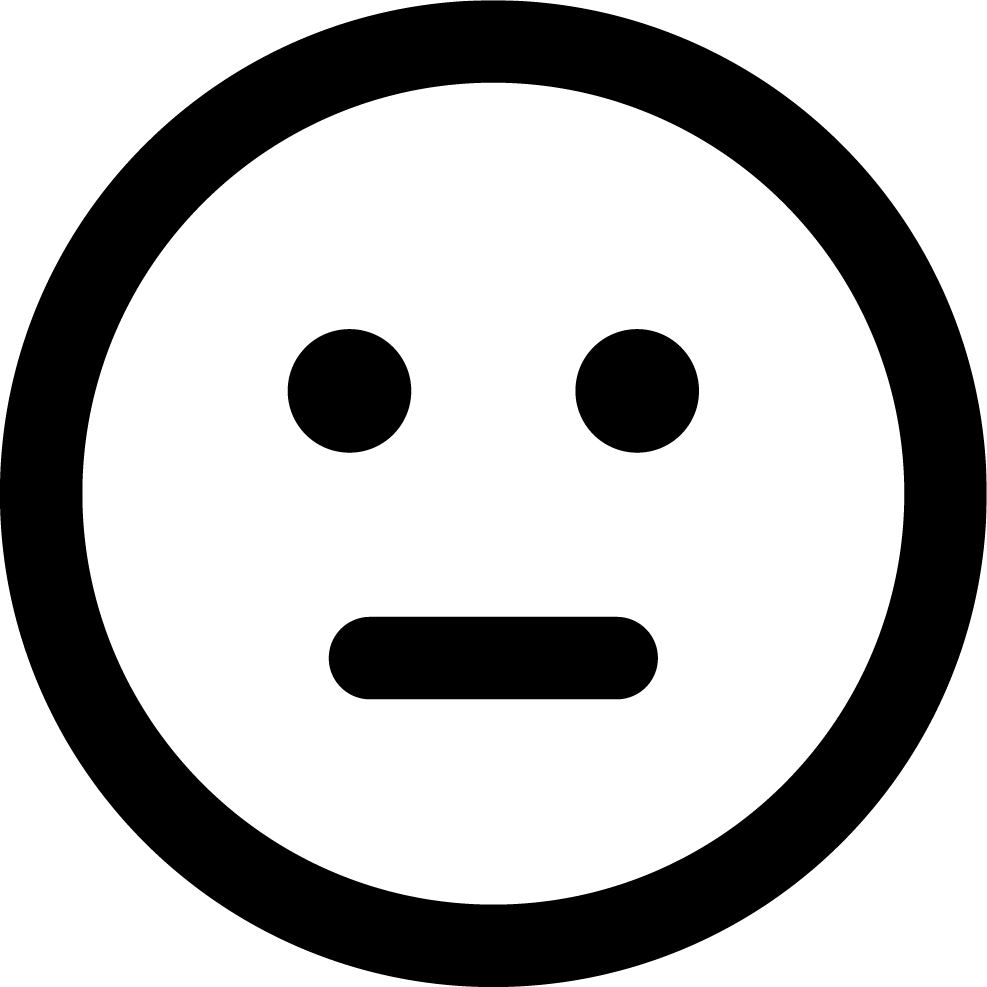}%
 \endgroup{} 
indicate responses that partially matched, and sad faces 
 \begingroup\normalfont
  \includegraphics[height=7pt]{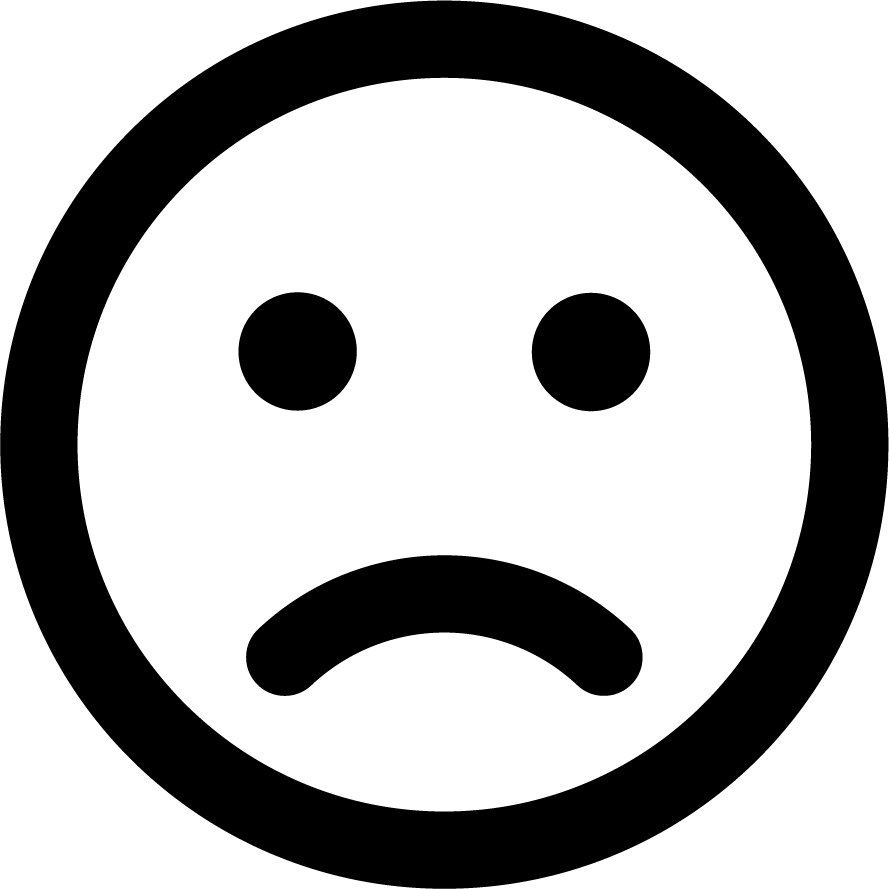}
\endgroup{} indicate responses that failed to match. The color code in the responses highlights discussions about the salient patterns in \textbf{\textit{\textcolor[RGB]{62,181,74}{data}}}, giving \textbf{\textit{\textcolor[RGB]{157,81,160}{design critiques}}}, or mentioning \textbf{\textit{\textcolor[RGB]{65,129,195}{statistical tasks}}} people observed.}
\label{fig:teaser}
\end{teaserfigure}

\maketitle

\section{Introduction}
\label{sec-intro}
Information visualizations help people extract meaningful analytical insights from data. For example, visualizations help people make sense of epidemiological data about COVID-19~\cite{kaul2020rapidly, borland2021enabling} or make predictions about natural disasters~\cite{cheong2016evaluating, preston2019uncertainty, ruginski2016non, millet2020hurricane}.
Graphical perception experiments measure the effectiveness of visualization designs~\cite{quadri2021survey}, \add{but the prevailing paradigm for graphical perception studies} focuses on low-level tasks~\cite{amar2005low}, typically measuring people's abilities to estimate individual, prespecified statistical quantities. For example, 
such studies might ask people to \textit{``estimate the correlation between IMDB rating and gross movie sales.''} (see \autoref{fig:stimulus_sc}(b)). The cuing used in these studies (i.e., instructing people to estimate a given statistic) may direct people's attention to statistics or patterns that are otherwise not immediately obvious~\cite{xiong2019curse}. 
\emph{High-level comprehension}, in contrast, describes the overall knowledge a viewer intuitively gains about the data without explicit cuing or guidance. Such comprehension reflects the salient statistics and patterns that emerge organically from a particular combination of data and design. While cued, low-level tasks tell if people \textit{can} perform a given task, understanding the data people comprehend in a visualization informs designers as to whether people \textit{will} perform that task.

We more holistically explore visualization interpretation to examine the alignment between designers' communicative goals and what their audience sees in a visualization, which we refer to as their \textit{comprehension}.
Understanding 
people's comprehension 
helps designers predict whether their objectives are reflected in how people interpret their visualizations~\cite{szafir2023visualization}. 
However, existing design guidelines are typically derived from either expert experience or explicitly-cued experiments looking \add{primarily} at low-level tasks \add{like correlation or value estimation}~\cite{kosara2016empire}. For example, the classical graphical perception paradigm found in experiments like Cleveland \& McGill~\cite{cleveland1984graphical}
measures visualization effectiveness as the ability to estimate specific statistical quantities. While people may be able to efficiently complete a task when cued, that pattern may not stand out when the same people encounter a similar visualization in the wild. 
Broadening visualization evaluation beyond low-level tasks can offer valuable insights into their effectiveness, considering factors like 
{knowledge acquisition} or key messages~\cite{burns2020evaluate}.
For example, 
past work has assessed people's understanding of a graph using six levels of knowledge acquisition~\cite{burns2020evaluate}, 
offered frameworks for formulating communication intents as learning objectives~\cite{adar2020communicative}, and established that
interpreting graphs requires more than just visual properties of visualizations~\cite{stokes2022striking}. 
We draw on this tradition \cite{adar2020communicative,burns2020evaluate} to examine a complementary perspective to classical graphical perception paradigms, focusing on how capturing the patterns visualizations intuitively communicate in data aligns with people's expectations about data communication from design practice and past experiments.

We conducted a qualitative study to investigate the high-level patterns people naturally see when they encounter a visualization without a guiding task. This study is a preliminary investigation 
that aims to 
reconcile what experiments \add{and guidelines} say graphs are "good" at and what people 
comprehend \add{in visualized data.} 
This study provides an alternative lens on the classical graphical perception paradigm by 
approaching the same concept (i.e., what statistical estimates do visualizations afford) from a bottom-up perspective, emphasizing statistical concepts in comprehension (i.e., what estimates occur without any statistical task framing) and using the graph's original intention as a lens to connect to target tasks.
%
People described the patterns they saw and questions they could address
in variations of three common graph types: scatterplots, bar charts, and line graphs (see \autoref{fig:teaser}). 
%
Our stimuli 
reconstructed visualizations from popular media sources to reflect best practices across a number of design dimensions (see Section \ref{sec-stimuli}). 
We drew inspiration for our graphs from extensively studied visualization types, commonly encountered graphs in daily life and real-world statistical datasets.
Verbal and textual responses from participants were coded using axial coding to extract patterns with respect to their \textit{alignment with stated objectives} (as inferred from each source graph's accompanying text) and to identify salient properties of the \textit{data}, \textit{design}, and \textit{statistical quantities} that shaped people's comprehension.

The patterns people reported seeing in a visualization failed to fully match stated intentions in  59\% of tested visualizations. We found that \textit{a visualization's communication goals do not always align with the knowledge people draw from a graph} even when following best practices (Section \ref{sec-analysis_intention}).
 %
People's interpretations of a given visualization varied with both the features of the visualization itself (e.g., the number of subgraphs, amount of data encoded, labels and units of measurements, visual complexity) and people's individual backgrounds. 
%
While chart type was a strong predictor of graph comprehension, 
\textit{chart type was not sufficient to predict the information people extract from a graph} (Section \ref{sec-analysis_design}).
Guidelines built on insights 
from 
traditional graphical perception experiments did not fully 
capture people's high-level comprehension (Section \ref{sec-analysis_Task}).

\add{These findings confirm prior theoretical and empirical observations in communicative visualizations~\cite{hullman2011visualization,adar2020communicative} and visualization sensemaking~\cite{stokes2022striking,franconeri2021science,burns2020evaluate} calling for more comprehensive approaches to visualization evaluation. 
We cannot fully understand 
how people derive insights from graphs 
exclusively by 
experiments using cued tasks. 
This confirmation reinforces the need for a range of diverse methodological paradigms in visualization evaluation. }
We need to simultaneously understand the precision and salience of different analytical tasks in visualization design. 
The guidelines generated by combining such top-down \add{(i.e., low-level statistical studies) and bottom-up (i.e., high-level comprehension and other forms of cognitive understanding)} aspects of visualization interpretation can help designers optimize visualizations to rapidly and efficiently communicate a range of target patterns.

\noindent \textbf{Contributions.} Our primary contributions are:
 1) a study eliciting the properties people intuitively see in different visualization designs,
 2) a preliminary analysis of how high-level visual comprehension aligns with or contradicts design guidelines from isolated low-level task studies, and 
 3) 
 insight
 into how data type, complexity, composition, and design influence the patterns people extract from visualizations.
Our results indicate a need for general design guidelines that better consider the knowledge people naturally extract from a visualization

\section{Background}
\label{sec-background}


Understanding what people see in visualizations is critical for helping designers create effective visualizations: they can predict whether their intended goals are actually reflected in the audience's interpretation~\cite{szafir2023visualization}.
%
%
Established guidelines for creating effective visualizations originate from experiences or are generalized from studies where individuals are explicitly directed to seek specific patterns or statistics within the visualizations. However, 
real-world scenarios do not typically provide cues or guidance for interpreting graphs encountered in the wild.
To inform our study, we draw on past work in understanding and characterizing insight and measuring graphical perception.


\subsection{Visualization Tasks \& Insight}
A well-established maxim in visualization states that ``the purpose of visualization is insight''~\cite{card1999readings}. 
In visualization, insight defines the knowledge people obtain from data and serves as a unit of discovery.
Visualization evaluation often aims to determine to what degree visualizations help people develop insight into their data~\cite{saraiya2005insight,saraiya2006insight,plaisant2007promoting}. However, characterizing insight is complex and requires more than estimating a single statistic. North describes insight along five separate dimensions: complex, deep, qualitative, unexpected, and relevant~\cite{north2006toward}. 
More recent definitions integrate ideas from cognition and neuroscience 
to clarify 
that the definition of insight should be broader in visualization and analytics~\cite{chang2009defining}
%
or integrate heuristics to characterize the broader value of visualization~\cite{wall2018heuristic}. While past works vary in how they measure insight, they all agree that insight is nuanced, complex, and key to effective visualization. 

Insight emerges as people use visualizations to conduct a series of \textit{tasks}, binding together patterns and statistics to make sense of data~\cite{russell1993cost,pirolli2005sensemaking}. Definitions of visualization tasks characterize the units by which people analyze data to build insights~\cite{amar2005low,brehmer2013multi}. For example, Brehmer and Munzner~\cite{brehmer2013multi} 
provide a hierarchical typology of tasks, exploring how smaller actions combine
to complete larger goals. Conventional visualization guidelines and design processes emphasize smaller, focused tasks (often the $what$ in Brehmer and Munzner's typology). However, designs constructed using these guidelines may not uniformly support each constituent task: design guidelines may conflict, or attributes of visualization may cause certain patterns to be more salient than others~\cite{kong2018frames}. \add{Adar \& Lee explored the perspective of the fundamental mismatch between designers' intended communication goals and the language they employ, approaching it through the lens of communicative visualization as a learning problem~\cite{adar2020communicative}.} 
Our work builds on this idea by investigating how the insights people build through visualizations align with those the visualization intends to communicate. 

\subsection{Graphical Perception \& Comprehension}


Many 
design guidelines are grounded in \textit{graphical perception}, the study of how people perceive specific information in visualizations~\cite{cleveland1984graphical, szafir2023visualization}. 
These studies typically focus on task effectiveness for specific, readily quantified tasks, such as assessing correlations across a range of designs~\cite{rensink2010perception, harrison2014ranking, kay2016beyond}, modeling how precisely people estimate statistics across visual channels~\cite{kim2018assessing, szafir2018modeling, smart2019measuring}, and measuring how different visualization techniques support a range of tasks~\cite{albers2014task, correll2012comparing, saket2018evaluating}.
See Quadri and Rosen~\cite{quadri2021survey} for a survey.

While significant research has evaluated how well people can extract specific statistical quantities, these studies explicitly cue people as to what patterns to look for, asking participants to answer direct questions such as, "\textit{identify the highest stock price in last decade}" (see \autoref{fig:teaser}). 
We draw on past work in graphical perception to understand the relationship between design guidelines and the natural patterns people comprehend in visualizations (i.e., the tasks they perform without any cues).

We investigate this relationship using scatterplots, bar graphs, and line graphs, which are the most widely used and highly studied graph types according to recent surveys~\cite{quadri2021survey}. Their effectiveness has been explored across a range of tasks, including judging values~\cite{gleicher2013perception, hong2021weighted, tseng2023evaluating} and exploring clusters~\cite{jeon2023clams, quadri2022automatic, quadri2020modeling} using scatterplots, finding extremum~\cite{waldner2019comparison, srinivasan2018s} and comparison~\cite{wun2016comparing, talbot2014four} with bar graphs, and estimating trends~\cite{best2006perceiving, correll2017regression, albers2014task} and value retrieval~\cite{saket2018task, heer2010crowdsourcing} with line graphs.

While these studies inform us of what common visualizations can do, they suffer from two key limitations: 1) they present simpler visualizations compared to most real-world applications, and 2) they provide people with a specific, well-defined goal, engaging a potentially different set of perceptual mechanisms than when encountering a visualization in the wild under less constrained conditions. Prior works demonstrated how the comparison between visualization could be difficult for people when graphs are complex, have more items, and larger size~\cite{gleicher2017considerations}. Our study aims to include complex graphs that resonate with real-world examples.
\add{Furthermore, people's interpretation is influenced by additional factors such as rhetorical techniques from other fields~\cite{hullman2011visualization} or individual background~\cite{hall2021professional, peck2019data, burns2023we}. Additionally, 
assessing the communicative goal of visualization requires more than just low-level tasks~\cite{adar2020communicative}, and relying solely on visual properties may not be sufficient to predict its interpretation~\cite{stokes2022striking}.}
We can compare open-ended responses summarizing people's high-level comprehension against design guidelines reflected by expressed design intents and results from these studies to begin to understand 
people's graph comprehension and scaffold future investigations.

Guidelines from cued studies may not 
be sufficient to guide design 
in part because there 
are two opposing ways people might attend to information in visualizations:
top-down and bottom-up~\cite{gibson1972theory,gregory1974concepts,connor2004visual}.
Top-down attentional processes are correlated with goal-driven attentional control (i.e., people attend to marks based on a given task or objective)~\cite{van2004bottom} whereas bottom-up processes correlate with stimulus-driven attentional control (i.e., people attend to marks based on the visual features of the marks)~\cite{awh2012top}. While most graphical perception studies reflect goal-driven processes, we argue that, in many cases, visualization comprehension relies on stimulus-driven processes (e.g., when exploring unfamiliar data or encountering a narrative visualization without specific guiding text). Zacks \& Tversky \cite{zacks1999bars} found that when asked to simply describe a graph, people attend to different features in different graphs and that these differences lead to different conclusions about data. 
Their results suggest that people attend to information in visualizations in two different ways due to top-down and bottom-up attentional processes.
However, their explorations focused on simple, two-point graphs. We extend these ideas to understand stimulus-driven comprehension in more conventional visualizations.  

We can understand specific design guidelines by comparing the patterns people comprehend against a 
visualization's
stated communication objectives. 
A visualization's abilities to achieve these objectives are likely mediated by a variety of factors such as visual encoding~\cite{quadri2021survey} and data distributions~\cite{kim2018assessing}. 
%
%
For example, people may attend to different patterns based on past knowledge~\cite{zacks1999bars, shah2011bar, xiong2019curse}. People's inferences about a graph may be swayed by social cues~\cite{kim2017data}, \add{rhetorical framing~\cite{hullman2011visualization}, audience background~\cite{peck2019data,grammel2010information,burns2023we}}, affect~\cite{lee2022affective}, additional text~\cite{stokes2022striking}, or \add{language describing intent~\cite{adar2020communicative}}.
Design guidelines grounded solely in graphical perception studies may fail to account for such factors. 
{For example, Stokes et al. \cite{stokes2022striking} leverage open-ended description tasks on line charts containing varying amounts of text, ranging from no text to a written paragraph with no visuals, 
and found that charts with text or statistical components led to 
more accurate takeaways than charts with elemental or encoded texts.

\add{Visualization design guidelines for general audiences should consider the characteristics of the potential audience, including 
diverse backgrounds~\cite{peck2019data,hall2021professional}, graphical literacy~\cite{franconeri2021science}, and experience levels~\cite{grammel2010information,burns2023we}. 
Such general communication must be especially cognizant of novices. A 'novice' is a person who is inexperienced or new to \textit{visualization} or \textit{visual analytics}. Past work on novices frequently focuses on more qualitative approaches to understanding visualization use, similar to our approach, to determine how people with limited experience may approach visualization~\cite{heer2008creation, grammel2010information}. 
For example, Grammel et al.\cite{grammel2010information} investigated how novices create visualizations using commercial visualization software, focusing on the author's perspective rather than the data consumer's 
to describe barriers novices face in the data exploration process. Burns et al.~\cite{burns2023we,burns2020evaluate} showed that even when considering novices as a population, visualizations often fail to adequately consider their role in evaluation. 
}
\add{While our work does not focus on novices, we draw on a similar body of methods to assess the relationship between visualization design and high-level graph comprehension, focusing on the conclusions people draw from data in practice and contextualizing those against design intentions reflecting conventional design guidelines.}


\section{Methodology}
\label{sec-methodology}


We conducted a qualitative experiment to characterize the patterns and statistics people comprehend in common visualizations when they encounter a visualization without a guiding task. 
Our experiment investigates two primary research questions: 
\begin{description}
    \item[{[RQ1]}\label{Research:RQ1}]
    \noindent\fcolorbox{black}{gray!10}{%
        \parbox{0.9\linewidth}{%
           \textit{Do the patterns people see in visualizations match the visualization's stated objective?}
        }%
    }
    \vspace{5px}
    \item[{[RQ2]}\label{Research:RQ2}]
    \noindent\fcolorbox{black}{gray!10}{%
        \parbox{0.9\linewidth}{%
           \textit{What types of patterns do people naturally see in common visualization designs?}
        }
    }
\end{description} 


We address these research questions in an empirical study using a combination of think-aloud and written free-response methods. The study asked participants to use various methods to describe the content of a sequence of graphs. These graphs were reconstructions of designs from the popular media \add{using available datasets (see Appendix B)}, where their stated communicative intention was extracted from the text of their accompanying article. 

%
We analyzed responses using axial coding and thematic analysis to identify preliminary patterns in the information that people intuit from visualizations. Our study characterized this intuitive data comprehension across three design dimensions: graph type (scatterplot, bar graph, and line graph), data type (single-class versus multi-class), and graph composition (juxtaposed versus non-juxtaposed). 


\subsection{Experimental Task}
\label{sec-experimental}

We designed two tasks to investigate whether graphs communicate the information they were designed to communicate (\textbf{RQ1}) and to elicit the specific types of patterns or statistics people notice in different visualization types (\textbf{RQ2}).
To tune our questions to ensure that they elicit the appropriate levels of comprehension, minimize potential bias from question wording, and establish a preliminary codebook, we conducted a pilot study with ten participants, asking them to describe five visualizations from the New York Times (c.f., Appendix A). Our formal study asked participants to use a graph to accomplish the following: 
\begin{description}
    \item[{[Description]} \label{Task:T1}] 
    \noindent\fcolorbox{black}{gray!10}{%
        \parbox{0.6\linewidth}{%
           \textit{Describe what do you see in the graph.}
        }
    }
    \vspace{5px}
    \item[{[Question]} \label{Task:T2}] 
    \noindent\fcolorbox{black}{gray!10}{%
        \parbox{0.75\linewidth}{%
           \textit{What questions would you be able to answer from this graph?}
        }
    }
   \end{description}

Approaching the task from two directions (as a description and a set of questions) reflects best practices in survey design to help overcome potential limitations from question framing \cite{krosnick2018questionnaire}. 
%
We used the \textbf{Description} task 
to 
elicit broader, contextualized observations such as insights~\cite{north2006toward}. 
%
%
The \textbf{Question} task implies tasks for which a ``correct'' answer could be inferred from the graph. This task is 
intended to prompt people to provide specific analytical tasks (e.g., estimate correlation) that a given visualization enables without cuing participants to individual statistics. 

%

\subsection{Study Design}
\label{sec-designdecision}
Visualizations in real-world scenarios encompass a vast array of designs, each potentially conveying distinct sets of statistical information. Building on the insights from the pilot study (c.f., Appendix A), our primary aim was to systematically design the stimuli in our study to provide consistency between stimuli while still reflecting the core design approaches in the original visualizations. 
To gain deeper insights into individuals' high-level comprehension, we need to understand how the interplay of design elements influences the information naturally communicated by diverse graphical representations. The influence of elements such as color, size, or mark shape may not be (and arguably often is not) separable in practice \cite{kim2018assessing,li2010size,szafir2018modeling}. Our approach instead provides a preliminary investigation of differences across aggregate design parameters in part to inform future studies that systematically vary individual components to understand their influence on comprehension. 
We chose to conduct the study on multiple visualization types---line graphs, bar charts, and scatterplots---rather than focusing on just one type to gain a broad preliminary understanding of how people interpret graphs, given the complex interactions we anticipate between design factors. These results help identify meaningful future investigations of specific variables in shifting comprehension. 

While there exists an abundance of empirical studies on commonly used graphs, there is a lack of guidance on high-level comprehension. \add{Moreover, earlier research has indicated that low-level tasks may not sufficiently address the true communication goals of visualization~\cite{adar2020communicative}.}
In this work, we focused on three graph types designed using best practices in visualization to communicate statistical information effectively. We aimed to draw from a diverse pool of visual designs to ensure a broad examination of the space grounded in real-world practices. We selected a range of frequently encountered and thoroughly studied graphs to provide preliminary comparisons across both heuristic and empirically grounded guidelines. We divided these factors into three categories describing the visualization type, data type (single versus multiclass), and composition (juxtaposed or non-juxtaposed; \autoref{fig:designabstract}).
This approach allowed us to explore the diversity of high-level comprehension across various visualization idioms and approaches.
We draw on expert-designed graphs from the popular media to help ensure best practices were enforced and provide context for extracting explicit design intentions. 
\begin{figure}[t]
\centering
\includegraphics[width=\columnwidth]{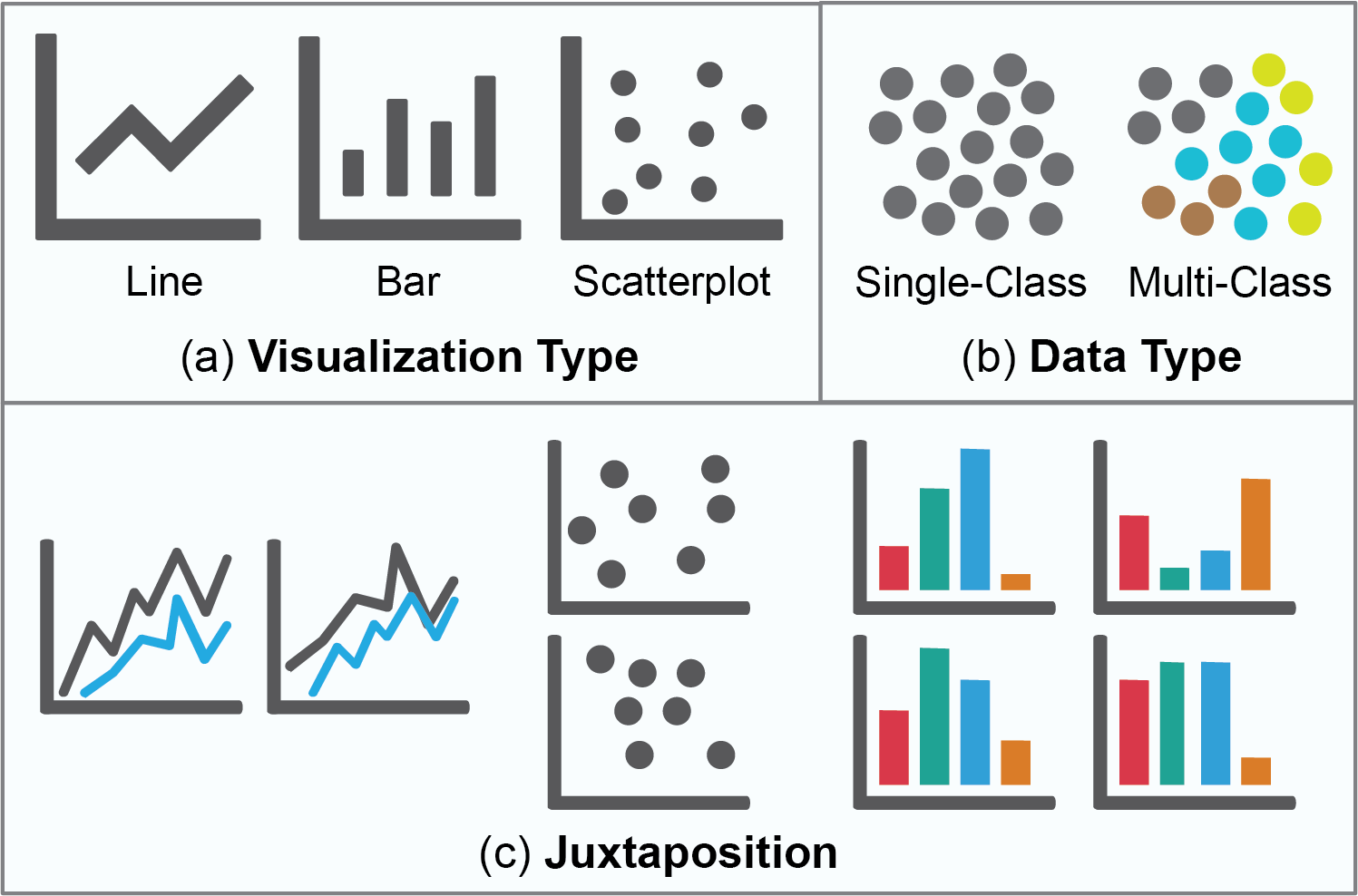}
\caption{The design dimensions (see \autoref{sec-stimulus}) used in our study: (a) visualization types, (b) data types, and (c) composition type (juxtaposed).}
\label{fig:designabstract}
\end{figure}

\subsubsection{Independent Variables}
\label{sec-stimulus}
Our study investigates visualizations that vary in \textit{visualization type, data type}, and \textit{graph composition} (see \autoref{fig:designabstract}). These three variables reflect general driving factors in visualization design: the core task people want to accomplish typically dictates the \textit{visualization type} used; the problem space dictates the \textit{data types}; and the combination of tasks and data (i.e., number and type of important attributes) determines \textit{composition}~\cite{munzner2014visualization}.

\vspace{3pt}\noindent
\textbf{Visualization Type:} We selected and tested 
\textit{scatterplots, line graphs,} and \textit{bar graphs}, as illustrated in \autoref{fig:designabstract} and \autoref{fig:stimuli_dimension}. These visualizations commonly appear in the media, and most participants will be familiar with their basic structure~\cite{borkin2013makes}. They are also among the most studied in classical graphical perception studies~\cite{quadri2021survey}, providing a robust baseline for comparison with results from cued experiments. 
%

\begin{figure*}[htbp]
\centering
\includegraphics[width=2\columnwidth]{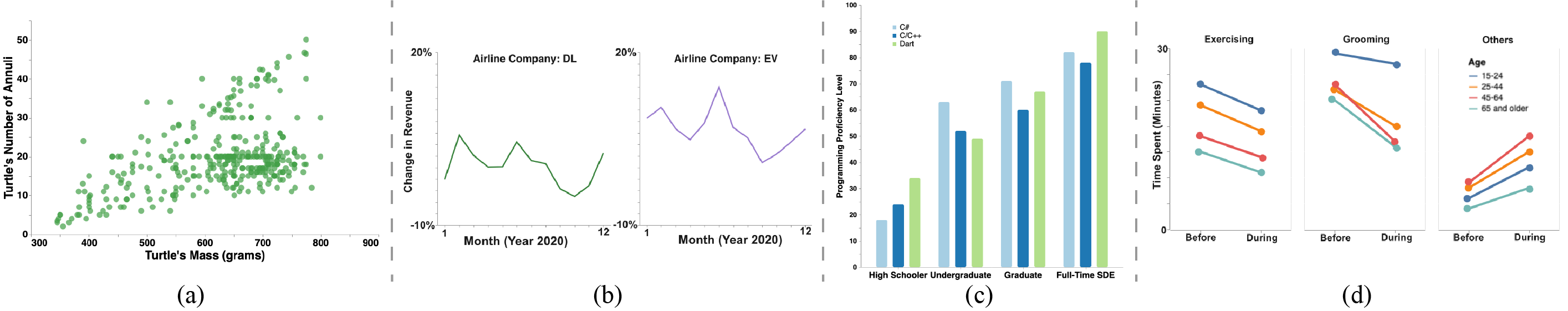}
\label{fig:stimuli_dimension}
\caption{The stimuli samples illustrating design dimensions (see \autoref{fig:designabstract} and \autoref{sec-stimulus}) used in our study, where visualizations in (a) use the \emph{Turtle} dataset~\cite{turtles}, dimensions: \emph{scatterplot}, \emph{single-class}, \emph{non-juxtaposed}. (b) A graph of \emph{Airline} dataset~\cite{airlines}, dimensions: \emph{line graph}, \emph{single-class}, \emph{juxtaposed}.
(c) A graph of \emph{ProgProficiency} dataset~\cite{programming}, dimensions: \emph{bar graph}, \emph{multi-class}, \emph{non-juxtaposed}.
(d) A graph of \emph{Activity-covid} dataset~\cite{mccarthy2021physical}, dimensions: \emph{line graph}, \emph{multi-class}, \emph{juxtaposed}.
}
\label{fig:stimuli_dimension}
\end{figure*}

\vspace{3pt}\noindent    
\textbf{Data Type:} We selected datasets (see Appendix B) to plot graphs with either single-class (no categorical data) or multi-class (data can be decomposed into categories) data (c.f., \autoref{fig:designabstract} and \autoref{fig:stimulus_mc}). Considering data type allowed us to look at graphs across both categorical and continuous encodings as well as to explore responses for more complex visualizations, which often communicate multiple points and have more complex communicative goals, revealing patterns in the ways different design choices interact with one another. 

\vspace{3pt}\noindent
\textbf{Composition Type:} Previous research has indicated that visual tasks 
often require comparing data across groups or dimensions~\cite{gleicher2017considerations, gleicher2011visual}. These visualizations often require interpreting data across multiple charts and synthesizing those interpretations to understand the collective message. We considered such graph compositions 
through either juxtaposed (side-by-side and up-down, see \autoref{fig:designabstract}, or non-juxtaposed \cite{javed2012exploring} designs. Single-class non-juxtaposed graphs consisted of a single graph, while single-class, juxtaposed graphs had multiple graphs, each showing a single class of data. Multi-class juxtaposed graphs had multiple graphs, each showing multiple classes of data, while multi-class non-juxtaposed graphs showed multiple classes of data on the same axes (i.e., superimposed). See \autoref{fig:stimuli_dimension} for examples. 

\subsubsection{Stimuli}
\label{sec-stimuli}

Visualizations in the wild encompass a wide range of designs; however, different visualizations communicate different sets of statistics. As we lack guidance on which visualization design elements shift the information a graph naturally communicates, we needed to draw from a set of designs that represents sufficient diversity to avoid confounds from isolated artifacts (e.g., outliers in a dataset, poor color choices, etc.).
We drew the inspiration from graphs in data journalism (i.e., the New York Times or government websites). These graphs (a) allowed us to explicitly understand their intended goals by mining target statistics from the accompanying text and (b) reflect known best practices for communicating this data. 
We collected a set of 60 graphs from these sources to reproduce (5 per unique setting of our independent variables). 
Our reproductions used the general design schema from the original graphics, but we replaced the data and associated labels and values with data from 42 stimulus datasets (selected at random) \add{to remove potential interpretive bias associated with data semantics~\cite{xiong2019curse,kong2019trust}.} To maintain ecological validity, 
we used real-world datasets employed in previous visualization studies. 
The datasets were drawn from various sources, including past visualization, HCI research, and real-world datasets. 

\add{We tried to mirror the design choices 
from the original graph as faithfully as possible while minimizing potential framing effects when adapting visualizations to the target datasets. Some adaptation was necessary to provide consistency across stimuli to avoid distracting participants (e.g., we used a consistent axis design). }
We 
mapped the data using 
shapes, 
colors, and data encoding channels (both continuous and categorical) from the original graphics. We maintained the aspect ratio, layout, and legend position. 
%
In the reconstruction process, we 
removed labels, titles, or captions 
(see \autoref{fig:reconstruction}).
%
\add{As we wanted to understand people's visual comprehension of graphs based on the visualization design and data alone, framing effects from titles~\cite{kong2019trust}, captions~\cite{cheng2022captions}, and labels~\cite{kong2019trust},  would have introduced a significant confound by cuing particular interpretations. 
For example, they may have provided information that guides readers towards a particular conclusion~\cite{rogers2022cues}}. 


%

We selected five graphs for every combination of 
visualization type (3) $\times$ data type (2) $\times$ composition (2), resulting in 60 stimuli.  \autoref{fig:designabstract} and \autoref{fig:stimuli_dimension} show examples of our \add{designed} stimuli \add{where a different dataset is used for the same design intent}. See the \href{https://osf.io/869ev/}{\textcolor[RGB]{0,0,255}{OSF Supplement}} for additional details and the full set of tested images and data sources.

\begin{figure}[b]
\centering
\includegraphics[width=1\columnwidth]{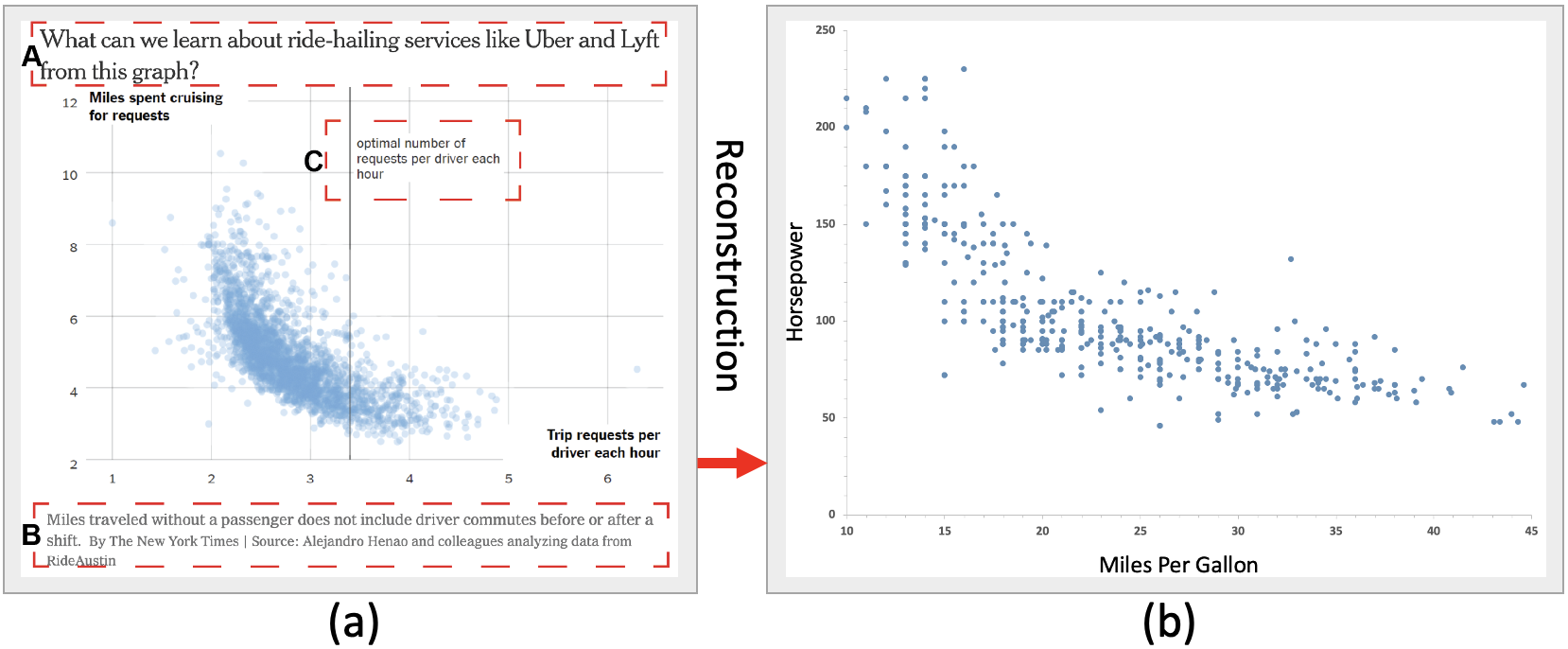}
\caption{Example figure in (a) illustrating a single-class non-juxtaposed scatterplot from \href{https://www.nytimes.com/2019/10/31/learning/whats-going-on-in-this-graph-nov-6-2019.html}{New York Times} showing negative correlations between two variables. To reconstruct this graph in (b), we removed additional text and labels (A \& B in (a)) and annotation (C in (a)) and plotted using another dataset~\cite{MPG_2017}. The stated objective is extracted from the text accompanied by the article.}
\label{fig:reconstruction}
\end{figure}
\subsubsection{Inferring 
Stated Objectives}
\label{sec-objective}
%
%

The intention behind each visualization in our corpus was inferred from the text accompanying each original source visualization. We first extracted the tasks associated with each source. We then mapped the visualization to the stimulus datasets and manually verified that the tasks remained appropriate for the resulting image (e.g., the intended tasks were still salient given the new data distribution and revised data semantics).  
For instance, in \autoref{fig:stimulus_sc} (a), the scatterplot was designed to visually represent sunny weather conditions. It achieved this by plotting the maximum temperature recorded on each day within a particular calendar year. The visualization's primary goal is to communicate maximum temperature patterns associated with sunny days and/or the distribution of temperature over time. 
The \href{https://osf.io/869ev/}{\textcolor[RGB]{0,0,255}{OSF Supplement}} contains the stated objectives for all 60 tested stimuli.

\subsubsection{Procedure}
Participants began by providing informed consent. The experimenter then explained the basic task (to describe each visualization and questions the visualization could address) and directed participants to a web application where they completed the formal trials. We did not include any tutorial examples to avoid potential bias from priming. 

Participants completed the two target tasks (\autoref{sec-experimental}) for a sequence of twelve visualizations, providing their responses to each question in a textbox. Each participant saw one chart from each combination of visualization type, data type, and composition---12 total visualizations. Visualizations were selected at random from the  
corresponding rendered stimuli for each combination of independent variables and presented serially in a random order. We encouraged participants to verbalize their reasoning processes to collect additional insight into strategies, points of confusion, and salient information that was not reported in the open-ended response text. 
%
\add{ To ensure task understanding and broaden our dataset, we 
instructed participants
to be as thorough as possible in their responses, both in the textbox and verbally, and allow participants as much time as they wanted to complete the study. The experimenter was present in the room to address any questions participants had at any point during the study. 
}
They finished the study by providing basic demographic information. 


Participants completed the study in 30-45 minutes. We recorded all participant study sessions for later transcription and analysis.

\subsection{Participant Recruitment}

We recruited 24 participants (18-56 years of age; 18 female, 6 male) from a University campus who participated either in-person (21 people) or via Zoom (3 people). Five identified as working professionals, and nineteen as students. \add{We did not intentionally vary or recruit participants with diverse levels of visualization literacy and novice expertise.} More details on the participants' demographics are in Appendix B. Potentially identifying information has been redacted from the quotes presented in this paper to protect participant anonymity in accordance with our IRB protocol.

\subsection{Data Analysis}
\label{sec-comprehension}
We used axial coding to analyze the natural language inputs from both the text responses and transcribed verbal utterances from the think-aloud protocol using the codebook derived in our pilot study (c.f., Appendix B).
Two coders each coded responses for 14 of the 24 total participants. 
%
They discussed the codes to refine the codebook after four transcripts. A summary of the codes is available in \autoref{tab:resultcoding} and an illustration in Appendix B. They both coded two overlapping sessions to verify interrater reliability. These sessions had a high overall agreement between the coders ($\kappa = 0.81$). In total, we have 288 responses collected from 24 participants. 
We coded each response for its \textit{comprehension match}, \textit{\textcolor[RGB]{65,129,195}{statistical tasks}}, \textit{\textcolor[RGB]{157,81,160}{design critiques}}, and \textit{\textcolor[RGB]{62,181,74}{data critiques}}.

\begin{figure*}[htbp]

\centering
\includegraphics[width=1.95\columnwidth]{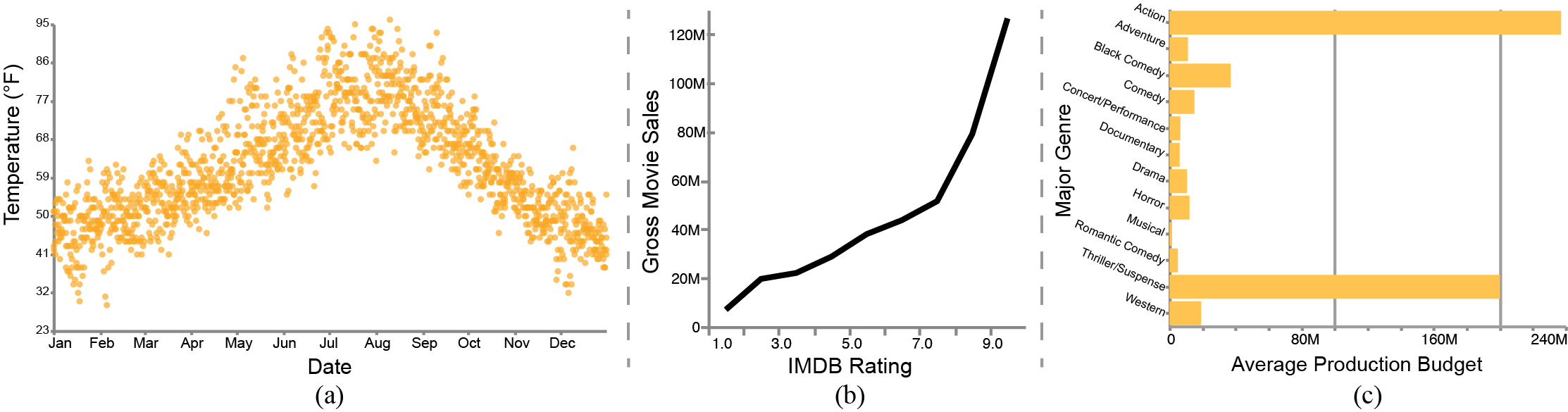}

\caption{Examples of Single-Class stimuli in our study. (a) is a Single-Class scatterplot with \emph{Sunny day} dataset~\cite{sunnydays}, (b) is a Single-Class line graph with \emph{IMDB} dataset~\cite{imdb}, and (c) is a Single-Class bar graph with \emph{Budget} dataset~\cite{movie}.}
\label{fig:stimulus_sc}
\end{figure*}

\textbf{Comprehension Match} 
quantifies the extent to which a participant's responses align with the stated objectives, as inferred from the accompanying text of the source graph (Section \ref{sec-objective}) and the participant's natural language descriptions and verbal responses. This measure is contingent on the degree of granularity in the statistical information discerned by participants from the presented graphs. This metric helps evaluate whether participants accurately perceive and convey the statistical information or insights that the visualization was designed to communicate. The alignment of the reported tasks and stated objectives provides insight into how intuitively the design communicates target information. 
\add{Codes of the intentions and responses focused on statistical concepts rather than semantic concepts to give consistency across visualizations by translating takeaway messages laden with dataset-specific semantics to objectives that hold across datasets (e.g., ``proficiency increases'' becomes ``increasing trend''). During data coding, we focused on aligning abstract statistical concepts and on word-to-word matching.}
%



Matches aligned at one of four thresholds: 

\begin{enumerate}
    \item Complete Match (CM): Specific statistics and patterns in the response matched the stated objective. A complete match indicates a strong alignment between participant responses and design objectives. In such cases, participants have effectively comprehended the key statistical information and insights intended by the original graph.

    \item General Match (GM): Participant's responses indicated the general knowledge they took from the graph aligned with the design's stated objective, but not specific statistics or patterns. 
    While there might be some minor differences or details that participants did not capture precisely, the core message and insights align well. For example, a \autoref{fig:stimulus_sc}(a) intends to \textit{document the pattern in the days where sunny weather occurs and the maximum temperature of each recorded day in a city from January to May and from August to December in 2012}.[P03] response matched with intent-- \textit{"Changes in temperature across the year and particular weather characteristics of each day (e.g., sun). The temperature starts low in January, goes up until August, and goes down from there."}.
    
    \item Partial Match (PM): Participant's responses partially matched the stated objective. They missed main statistical quantities or general knowledge about the data. For example, [P07] responded-- \textit{"the graphs describe the temperature of different months in 2012 under the different weather"} on \autoref{fig:stimulus_sc}(a) and missed statistical quantities on how the temperature follows a pattern- starts low (around 50 F) in January, goes up (above 85 F) until August, and goes down September onwards.
    
    \item No Match (NM): Participant's responses do not match with the stated objective and missed critical information entirely. Participants have not effectively comprehended or articulated the information the graph was designed to convey. A no-match implies that participants may have misunderstood or missed intended information.
\end{enumerate}

 \textbf{\textcolor[RGB]{65,129,195}{Statistical Tasks}} encode the patterns and statistics participants provide in their responses. We employed Amar et al.'s low-level task taxonomy~\cite{amar2005low} to code these tasks as it provides sufficient coverage of the tasks in our stated objectives and participant responses. Codes included \textit{Retrieve Value}, \textit{Filter}, \textit{Compute Derived Value}, \textit{Find Extremum}, \textit{Sort}, \textit{Determine Range}, \textit{Characterize Distribution},  \textit{Find Anomalies}, \textit{Cluster}, \textit{Correlate}, and \textit{Compare}.

\textbf{\textcolor[RGB]{157,81,160}{Design Critiques}} identify feedback provided by participants about how a particular design choice assisted or hindered their
comprehension of a visualization. We did not ask participants to provide critiques directly, but documented critiques that arose naturally in the responses or think-aloud.

\textbf{\textcolor[RGB]{62,181,74}{Data Critiques}} identify places where the data itself inhibited high-level visual comprehension. This included comments on missing context, needing additional dimensions, or failure to understand the data directly (e.g., jargon or acronyms). While again we did not prompt for these values, they arose in many participants' responses and provided additional insight into gaps in data comprehension and opportunities for improved visualization design.

\begin{figure}[b]
\centering
\includegraphics[width=\columnwidth]{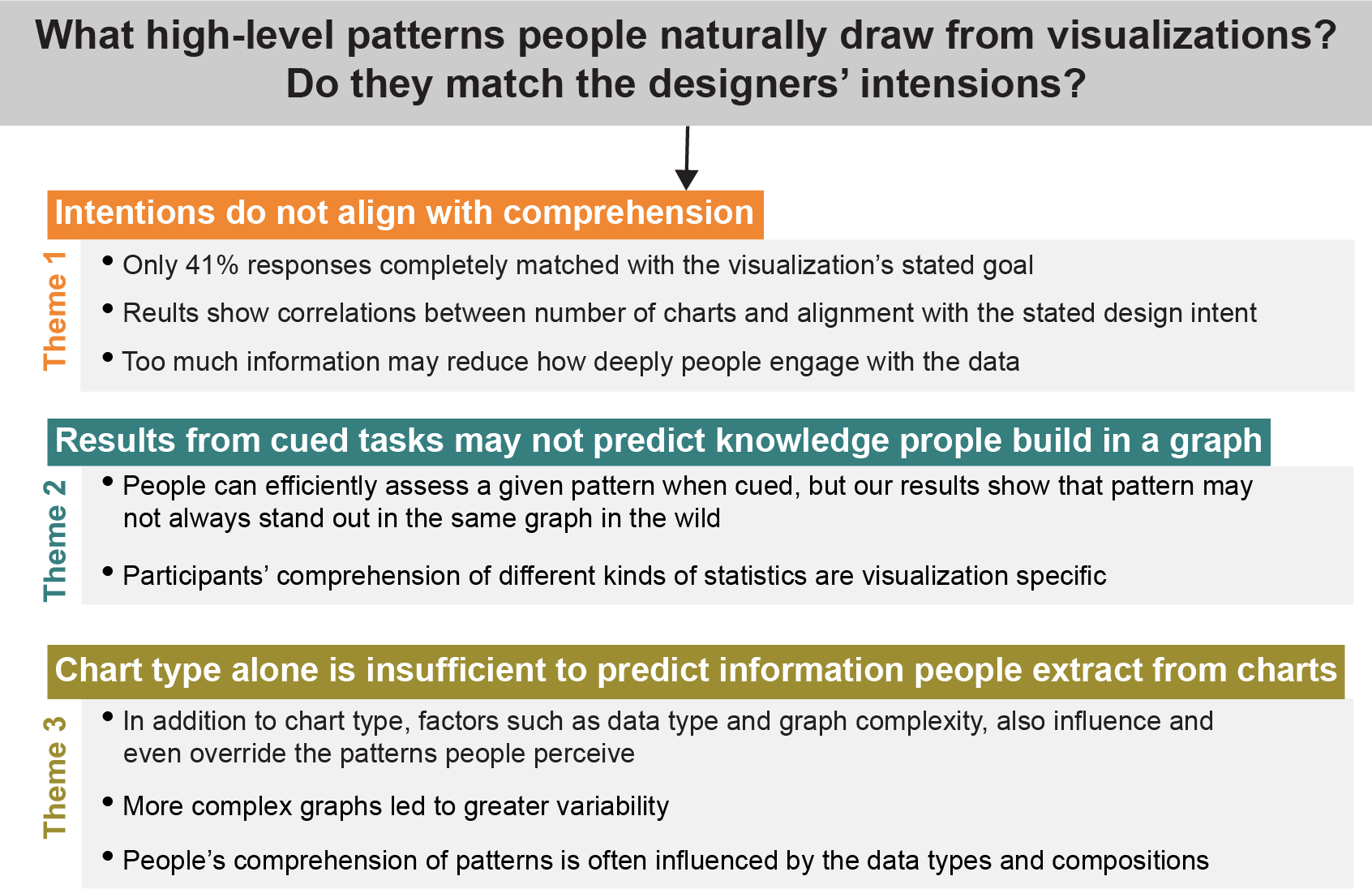}
\caption{\add{
A roadmap of the results from our study. We group our findings into three themes--- Theme 1 (see Section \ref{sec-analysis_intention}, Theme 2 (see Section \ref{sec-analysis_Task}), and Theme 3 (see Section \ref{sec-analysis_design}).
}}
\label{fig:roadmap}
\end{figure}

\begin{figure*}[htbp]
\centering
\includegraphics[width=1.95\columnwidth]{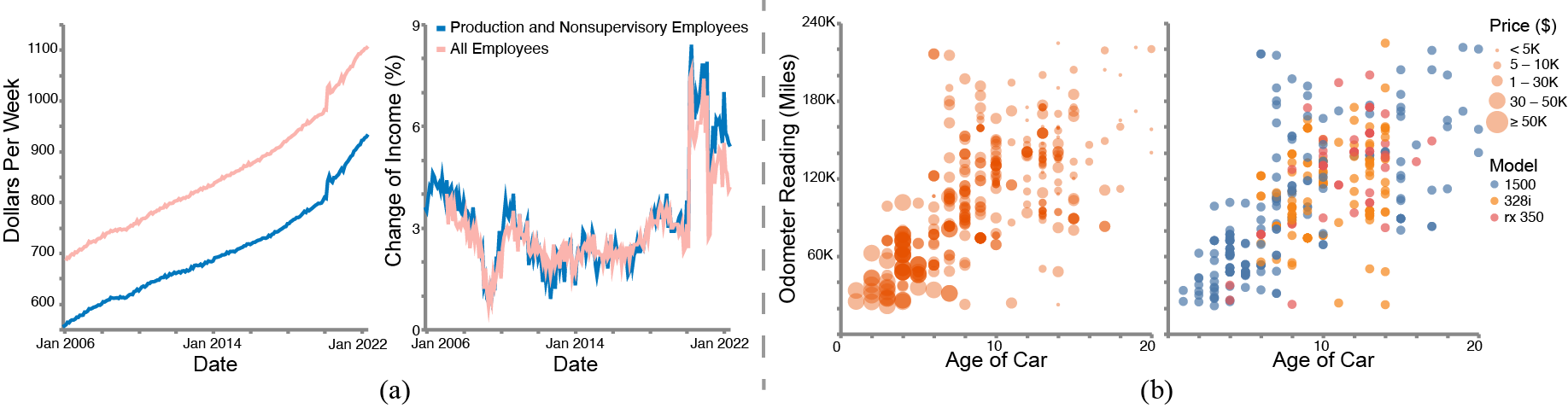}

\caption{Examples of Multi-Class stimuli in our study. (a) is a Multi-Class Juxtaposed line graph with \emph{Income} dataset~\cite{kohavi1996scaling}, and (b) is a Multi-Class Juxtaposed scatterplot with \emph{Car models} dataset~\cite{carmodels}.}
\label{fig:stimulus_mc}
\end{figure*}

\section{Thematic Analysis and Key Findings}
\label{sec-analysis}

A key motivation of this work is to characterize visualization design as a function of the patterns people perceive unprompted from a graph. In other words, do people see the information a visualization is designed to communicate? We analyzed participants' response data using thematic analysis and found three overarching themes in order to understand the need for visual comprehension as a metric (see \autoref{fig:roadmap} for the roadmap of our thematic analysis): 

\noindent   \textbf{Theme 1}:  \textcolor{orange}{\textbf{Intention does not align with comprehension.}} Stated communication goals 
   did not always align with the knowledge people drew from a visualization. Only 41\% responses completely matched with the visualization's stated goal even though our study's graphs followed the design schema and matched recommendations from prior work. This result echoes calls made in more comprehension-oriented studies \add{(e.g., \cite{burns2020evaluate,adar2020communicative})} to understand better the connection between design guidelines in theory and their efficacy in practice. \add{Moreover, the stated designs' intent may not always align with people's comprehension as comprehension is influenced by multiple factors, including visualization design, individuals' background, graphical literacy, and supplemental information.}
  
  \noindent  \textbf{Theme 2}:  \textcolor{teal}{\textbf{Results from cued tasks alone may not predict knowledge people build in a graph.}}
   The prevalent model in visualization research defines visualization effectiveness as the ability of the average user to complete a given task~\cite{quadri2021survey}. 
    This assumption has limited ecological validity: people may be able to efficiently assess a given pattern when cued, but that pattern may not stand out in the same graph in the wild. 
    If someone \textit{can} use a given graph to accomplish a directed task, it does not mean that they \textit{will}. Even though prior work showed certain visualization types (e.g., scatterplot, bar, and line graphs) effectively convey various statistics and patterns (e.g., correlation, distribution, cluster, anomalies), less than 50\% of participants' responses included such patterns. \add{This observation supports the argument made in past work~\cite{adar2020communicative, kosara2016empire, north2006toward}, emphasizing that low-level tasks alone are not sufficient to address the real communication goals of graphs.}
    Our observations suggest guidelines built on insights from cued experiments do not fully capture people's high-level comprehension. \add{Instead, guidelines generated by combining low-level statistical studies and high-level comprehension may help designers optimize visualizations to communicate a range of salient patterns efficiently.}
    
  \noindent  \textbf{Theme 3}: \textcolor{olive}{\textbf{Chart type alone is not sufficient to predict the information people extract from a visualization}}
   Prior work in visualization effectiveness and corresponding tools like chart choosers often match visualization tasks to chart types~\cite{saket2018task, quadri2020modeling, wang2024empirical, kim2018assessing, mylavarapu2019ranked}. We observed that chart type tended to dictate the patterns people commonly reported (e.g., correlation in a scatterplot, trend in line graphs, and identifying discrete and maxima in bar graphs). However, data type and graph complexity (e.g., composition) also influence, and can even override, the patterns people perceive. We observed many mismatches between participants' responses and the visualizations' stated goals when graphs encoded multi-class data types using juxtaposed compositions (\autoref{fig:stimulus_mc}). \add{These observations validate 
   past findings that relying solely on chart types or visual properties is insufficient to predict how the communication goal will be interpreted ~\cite{stokes2022striking,cheng2022captions,hullman2011visualization}.}

Our thematic analyses also allowed us to investigate various design and data critiques provided by participants
(see Section \ref{sec-analysis_design_data}).  
\autoref{tab:resultcoding} summarizes comprehension matches for all 288 collected responses as a function of our independent variables (graph type, data type, and graph composition).

 \subsection{Theme 1: \textcolor{orange}{Intentions Do Not Align with Comprehension}}
\label{sec-analysis_intention}

%
%
We evaluated the \textit{comprehension match} (see Section \ref{sec-comprehension}) between the descriptions provided in the participant responses against the stated objectives of the original source graph (see Section \ref{sec-objective}). A complete match means the participant's description and questions identify all of the objectives inferred from the original source graph and may include additional information as well. Only  41\% (117/288) of responses completely matched the designers' objectives.  Additionally, 77\% (92/117) of those responses were provided by the 10 participants whose provided complete or general matches 
for at least 10 of their 12 stimuli (see Appendix B). 
This discrepancy implies that \textbf{an expert-crafted visualization's intended communication goals \add{may} not always align with the insights and salient patterns people intuitively extract from that visualization.}
The strength of this alignment varied by graph complexity, with multiclass non-juxtaposed line charts (33\% of responses were complete matches) and scatterplots (46\% complete matches) and multiclass juxtaposed line charts (37\%) and scatterplots (33\%) seldom matching the target objectives. 
We also found a correlation between people's comprehension matches and their self-reported familiarity with graphs and background, which we discuss in Section \ref{sec-perparticipant}.

The mismatch between reported descriptions and stated intentions signals a need for better guidelines: we assume that professional venues follow best practices for data communication and selected our source visualizations such that they followed this assertion. 
We can examine participant responses to investigate where discrepancies arise between a visualization's intended message and the knowledge people extract in practice. 
In the following discussion, we report general patterns but focus specific examples on three graphs---\autoref{fig:teaser}, \autoref{fig:stimuli_dimension}, \autoref{fig:stimulus_mc}, and \autoref{fig:stimulus_sc}---but additional details are available in Appendix B and \href{https://osf.io/869ev/}{\textcolor[RGB]{0,0,255}{OSF Supplement}}.

%
%
The alignment between intention and viewer comprehension is primarily influenced by the visualization design and the complexity of the graph, as summarized in \autoref{tab:resultcoding}. 
People commonly reported specific statistics for certain chart types (e.g., correlation in scatterplots, trend in line graphs, and identifying discrete values and maxima in bar graphs). However,  data type and graph composition also influence, and can even override, these patterns (see Section \ref{sec-analysis_design}).

Participants' data interpretation and comprehension varied with the visualization type used. Single-class scatterplot responses aligned with the inferred intention in 57\% of cases, while single-class bar graphs (65\%) and line graphs (68\%) had higher alignment. People tended to quickly identify the peaks and valleys and changes over time in line graphs, maximum values in bar graphs, and associations and correlations in scatterplots. 

In other cases, participants' responses significantly deviated from the intended communication goal
and focused on what they can read on graphs.
For example, for \autoref{fig:stimulus_mc}(a), which intended to \textit{show positive correlation and compare two types of employees’ average weekly incomes and their growth rate over 16 years}, they read the line graph using label and legends, identified `change' in income, noted that both graphs follow a similar trend, and attended to the highest values. One participant described the visualization as ``\textit{an increase in how many dollars are earned per week between earnings of production and non-supervisory employees and earnings of all employees} [P01]''. 
At times, the misalignment resulted from a lack of familiarity: ``\textit{I don’t know the scatterplot, [I'm] confused with the graph and terminology mentioned in the labels. But I am still understanding the graph about architecture.} [P07]'' (see  Fig.11 in \href{https://osf.io/869ev/}{\textcolor[RGB]{0,0,255}{OSF Supplement}}).

We found a \textbf{correlation between the number of charts and alignment with the stated design intent}. Specifically, fewer charts led to higher alignment. Non-juxtaposed charts tended to support higher alignment than other kinds of composite charts---scatterplot (23/48), line graph (20/48), and bar graph (19/48). Additionally, single-class juxtaposed compositions for both line (12/24) and bar graphs (13/24) achieved their intended goals better than composite graphs. For example, in the single-class, non-juxtaposed scatterplot in \autoref{fig:stimuli_dimension}(a), participants identified the intended positive correlation 
in 90\% of cases compared to 33\% in single-class juxtaposed scatterplots.

People frequently failed to find any relevant patterns in visualizations with multiple categories represented using multiple encodings (e.g., \autoref{fig:teaser}(a), 2/3, and \autoref{fig:stimulus_mc}(b), 2/5) where position, size, and color encode different data aspects, supporting the findings from Gleicher~\cite{gleicher2017considerations} that tasks grow difficult with increases in complexity. One participant mentioned---``\textit{I could understand bar graphs and simple line graphs well but not scatterplots or graphs involving multiple categories.} [P16]'' We observed more than 30\% of responses failed to identify any alignment with the stated objectives (37\% in scatterplots, 32\% in line graphs, and 34\% in bar graphs).

Several misalignments arose when participants found the graphs difficult to interpret. These instances largely occurred in complex visualizations, most notably juxtaposed graphs and multiclass data (juxtaposed scatterplots, 16/48 responses failed to match; juxtaposed line graphs, 15/48; juxtaposed bar graphs, 14/48; multiclass line graphs, 9/24; and multiclass scatterplots, 4/24). One participant noted that ``\textit{It is easier to read bar charts and simple line graphs but not scatterplots or graphs involving multiple categories} [P13]". 

In these cases, people tended to describe graphs using surface-level information rather than digging into statistical patterns in the data. 
For example, participants read and described what a graph represents through legends and axis labels (scatterplots 25/96, line graphs 29/96, bar graphs 24/96). 
\autoref{fig:teaser}(a)
is intended to show the distribution between patient's age and their BMI, support comparisons across in four different regions, and build relations between the distributions of four variables---age, BMI, region and insurance charges---across regions. Two of the three participants who interpreted this 
visualization focused on surface-level data, describing the data in the graph (BMI based on the region, age, and gender) but not mentioning any patterns in the data. One described it as ``\textit{looking at patient age, BMI and comparing gender and geographic location.} [P24]'' while the other said `` \textit{This graph shows client BMI vs. client age in diff regions of the US while differentiating between males and females.} [P05]''  
Only one out of three participants' responses matched the stated intention described in the source graph---``\textit{The graph seems to be depicting BMI for females and males based on age and geographic location. There does not seem to be a clear trend based on the graphics depicted. BMI, age, and geographic location are not clearly correlated.} [P24]'' More complex messages, such as the absence of correlation, can be difficult to communicate. However, our results indicate \textbf{a trade-off between complexity and interpretation: too much information may reduce how deeply people engage with the data.}

\subsection{Theme 2: \textcolor{teal}{Results From Cued Tasks Alone May Not Predict Knowledge People Build in a Graph.}}
\label{sec-analysis_Task}

\label{sec-statistical}
\begin{figure}[t]
\centering
\includegraphics[width=0.8\columnwidth]{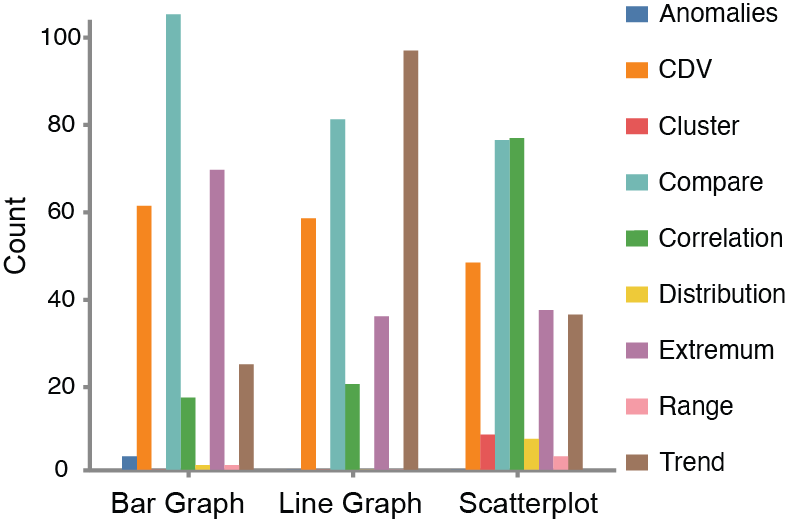}
\caption{Summary of participants' responses on \textcolor[RGB]{65,129,195}{statistical quantities} and \textcolor[RGB]{65,129,195}{patterns} in their comprehension. Some statistics and patterns are specific to a particular graph type, for example, extremum on bar graphs, trend on line graphs, and correlation on scatterplots. }
\label{fig:statistic}
\end{figure}

Visualization guidelines suggest what tasks people perform well with a given visualization. For example, scatterplots support clustering, characterizing distributions, and correlation; line graphs convey trends and temporal patterns; and bar charts communicate distributions, discrete value identification, and extremum (see Quadri \& Rosen~\cite{quadri2021survey} for a survey). However, these assertions about task effectiveness correspond to the ability of the average user to complete a given task when prompted. This assumption has limited ecological validity: \textbf{people can efficiently assess a given pattern when cued, but our results show that pattern may not always stand out in the same graph in the wild}. 
%
\begin{table*}[htbp]
\centering
\caption{Summary of participant responses alignment with designer's intention per chart type. 
SC: single-class; MC: multi-class; SC-J: single-class juxtaposed/overlayed; MC-J: multi-class juxtaposed/overlayed. To explore more on comprehension match, see per participant analysis in Appendix B.}
\label{tab:resultcoding}
\scalebox{0.95}{
    \begin{tabular}{ |c|cccc|c|cccc|c|cccc|c|c|  }
     \hline
     \multirow{2}{*}{Response Coding} & \multicolumn{5}{|c|}{Scatterplot} & \multicolumn{5}{|c|}{Line Graph} & \multicolumn{5}{|c|}{Bar Graph} & \multirow{2}{*}{All}  \\
     \cline{2-16}
     ~ & SC & MC & SC-J & MC-J & All & SC & MC & SC-J & MC-J & All & SC & MC & SC-J & MC-J & All & ~  \\
     \hline
     Complete Match & \cellcolor{green!12}12  & \cellcolor{green!10}11 & \cellcolor{green!8}8 & \cellcolor{green!9}8 & \cellcolor{gray!39}39  & \cellcolor{blue!12}12 & \cellcolor{blue!8}8 & \cellcolor{blue!11}11 & \cellcolor{blue!9}9 & \cellcolor{gray!40}40  & \cellcolor{red!11}11  & \cellcolor{red!8}8  & \cellcolor{red!12}12  & \cellcolor{red!7}7 & \cellcolor{gray!38}38   & \cellcolor{gray!85}117  \\
     General Match & \cellcolor{green!3}3   & \cellcolor{green!3}3  & \cellcolor{green!1}1  & \cellcolor{green!3}2  & \cellcolor{gray!10}9   & \cellcolor{blue!3}3  & \cellcolor{blue!0}0  & \cellcolor{blue!1}1  & \cellcolor{blue!1}1  & \cellcolor{gray!5}5   & \cellcolor{red!3}3  & \cellcolor{red!5}5  & \cellcolor{red!1}1  & \cellcolor{red!3}3  & \cellcolor{gray!12}12   & \cellcolor{gray!27}26   \\
     Partial Match & \cellcolor{green!4}4  & \cellcolor{green!6}6  & \cellcolor{green!8}8  & \cellcolor{green!4}5  & \cellcolor{gray!22}23   & \cellcolor{blue!3}4  & \cellcolor{blue!6}7  & \cellcolor{blue!3}3  & \cellcolor{blue!8}8  & \cellcolor{gray!20}22   & \cellcolor{red!4}4  & \cellcolor{red!7}7 & \cellcolor{red!6}6  & \cellcolor{red!5}5  & \cellcolor{gray!22}22   & \cellcolor{gray!64}67   \\
     No Match & \cellcolor{green!5}5   & \cellcolor{green!4}4  & \cellcolor{green!7}7  & \cellcolor{green!6}9  & \cellcolor{gray!22}25  & \cellcolor{blue!6}5  & \cellcolor{blue!10}9  & \cellcolor{blue!9}9  & \cellcolor{blue!6}6  & \cellcolor{gray!31}29   & \cellcolor{red!6}6  & \cellcolor{red!4}4  & \cellcolor{red!5}5  & \cellcolor{red!9}9  & \cellcolor{gray!24}24   & \cellcolor{gray!77}78   \\
     \cline{1-17}
     All   & \cellcolor{gray!24}24  & \cellcolor{gray!24}24 & \cellcolor{gray!24}24 & \cellcolor{gray!24}24 & \cellcolor{gray!60}96  & \cellcolor{gray!24}24 & \cellcolor{gray!24}24 & \cellcolor{gray!24}24 & \cellcolor{gray!24}24 & \cellcolor{gray!60}96 & \cellcolor{gray!24}24 & \cellcolor{gray!24}24 & \cellcolor{gray!24}24 & \cellcolor{gray!24}24 & \cellcolor{gray!60}96 & \cellcolor{gray!100}288\\
     \hline
    \end{tabular}
}
\end{table*}

Our analysis showed that just because someone \textit{can} use a given graph to accomplish a directed task does not mean that they \textit{will} without guidance. Measuring performance using abstract statistical quantities may not tell designers whether their visualizations will likely achieve their goals. 
For example, people did not identify \textit{correlation}, \textit{data distribution}, or \textit{outliers} in the line graph in \autoref{fig:stimulus_mc}(a) 
even though line graphs effectively communicate these statistics~\cite{harrison2013influencing, albers2014task}. This section explores how our results confirm or override previous graphical perception experiments. \autoref{fig:statistic} summarizes our results.

Prior work demonstrated that people are good at estimating distributions~\cite{chan2013generalized, saket2018task}, clusters~\cite{sedlmair2012taxonomy, quadri2020modeling}, and correlations~\cite{harrison2013influencing, rensink2010perception} in scatterplots. In contrast to past work, participants did not mention correlations in their scatterplot descriptions for 65\% of scatterplots (reported 75 times on 216 tested graphs), clusters for 84\% (8 reported of 48),
and 
distributions in 97\% (7 reported of 240) of responses.
In several cases, participants focused on the meanings of individual points in a scatterplot rather than the relations between them: ``\textit{I am seeing a dot graph showing different models of cars with different price points measured by their odometer reading at a specific age. I can tell how many miles a specific car has based on age.} [P15]"; or ``\textit{key telling the price of the car and the model of the car. The age of the car and the odometer reading in miles.} [P23]'' for \autoref{fig:stimulus_mc}(b). Further, people did not discuss statistics such as extremum or computing derived values.

People frequently described \textit{trends} in line graphs (96 times); however, they seldom performed other tasks that line graphs are known to be effective for, such as characterizing distributions (no responses)~\cite{ gogolou2018comparing, albers2014task}, estimating correlation (20 responses)\cite{harrison2014ranking, kay2016beyond}, and detecting anomalies (no responses)~\cite{albers2014task, saket2018task}. 
Multi-line graphs further failed to communicate distribution (none), clusters (none), and correlation (3 responses). 
Participants tended to describe each individual graph rather than to draw relations between them---``\textit{There are two graphs. From the right one, I see their income change over the years. I am seeing two line graphs showing the average weekly earnings of 2 categories of employees over the period of Jan 2006 to Jan 2022 and their change of income over the previous year.} [P11]" on \autoref{fig:stimulus_mc} (a). 
This lack of synthesis suggests that people may need to be explicitly cued as to how to connect data or findings across visualizations. 
For example \autoref{fig:stimulus_mc}(a) aimed to communicate the correlation between earnings of supervisory and non-supervisory employees with the length of employment and their raises over time; however, participants predominantly described the graph according to local trends---``\textit{...change in the employees wage with time}" [P02, P08, P09, P11]---overlooking the comparisons the chart composition afforded.

People's responses generally confirmed prior findings for bar graphs on discrete value identification (in 60 of the responses)~\cite{xiong2019biased,waldner2019comparison, saket2018task}, extremum (67), and trend (25)~\cite{srinivasan2018s,ondov2018face}. However, participants seldom mentioned distribution (3) and anomalies (5), despite their known effectiveness \cite{saket2018task,srinivasan2018s}. Though we saw better 
correspondence between statistical tasks and known perceptual results in bar graphs, several cases did not fully align with previous findings~\cite{srinivasan2018s,saket2018task}.

Participants' \textit{comprehension of different kinds of statistics are visualization-specific}, see \autoref{fig:statistic}. As expected ''trend'' is 
heavily associated with line graphs (in 96 of the responses).
Data also seemed to play a role: participants focused on extremums at longer bar heights for bar graphs and spikes in line graphs. High correlation ($>$ 0.9) in scatterplots and line graphs and notably valleys and low peaks in line graphs attracted viewers' attention, resulting in higher association with corresponding statistical tasks.


\subsection{Theme 3: \textcolor{olive}{Chart Type Alone is Not Sufficient to Predict the Information People Extract From a Visualization}}
\label{sec-analysis_design}

Prior work in visualization effectiveness and corresponding tools like chart choosers often match visualization tasks to chart type~\cite{quadri2021survey}. As discussed in Section \ref{sec-analysis_intention}, chart type dictated the patterns people commonly reported (e.g., correlation in a scatterplot, trend in line graphs, and identifying discrete values and maxima in bar graphs). However, other factors, such as data type and graph complexity (e.g., composition), also influence, and even override, the patterns people perceive. 

We observed several effects of design dimensions beyond chart type on visualization interpretation, specifically grouped around chart complexity (as modeled by data type and composition), scaffolding, and individual differences amongst participants. More complex graphs (e.g., multi-class data and juxtaposed sub-graphs) led to greater variability in the people's identified patterns and lower overall alignment with the visualization's stated objective. 
Added chart scaffolding (e.g., annotations or legends) led to higher alignment, while data anomalies (e.g., outliers) tended to change the patterns people reported from a given graph type.

\subsubsection{Data Type \& Distribution} 
\label{sec:data_distro}
People's comprehension of patterns was 
often influenced by data type---single- and multi-class---and 
compositions. People more readily identified patterns in single-class graphs as compared to multi-class data. For example, people frequently identified intended statistics in \autoref{fig:stimulus_sc}(a) (distribution: 4/4), \autoref{fig:stimulus_sc}(b) (correlation: 2/3), and \autoref{fig:stimulus_sc}(c) (correlation: 4/5) (see Appendix B for the full list). People's alignment with a visualization's stated goals was significantly lower in multi-line graphs (partial + no-match $>$ 66\% on multi-class non-juxtaposed and 58\% on multi-class juxtaposed), scatterplots with multiple data dimensions (partial + no-match $>$ 41\% on multi-class non-juxtaposed and 58\% on multi-class juxtaposed), and multi-class bar graphs ( partial + no-match $>$ 46\% on multi-class non-juxtaposed and 58\% on multi-class juxtaposed).
 
 The observations provided in the descriptions of graphs with multiclass data were limited and tended to focus on correlation and trend while missing other patterns, such as the relations between two sub-graphs. 
For example, in the case of the multi-class juxtaposed line graph \autoref{fig:stimulus_mc}(a), 
only one of eight responses aligned with the stated communication goal of that graph---``\textit{I see changes in dollars (which I assume represents weekly income) of weekly earnings for all employees at an unspecified company versus earnings of employees that work in production or non-supervisory roles... It seems that production and non-supervisory employees tend to make less money per week compared to the employee pool, but salaries have changed at comparable rates.} [P09]". 

However, even single-class charts exhibited strong variance between participants. For example, in a single-class scatterplot (see Fig.1 in \href{https://osf.io/869ev/}{\textcolor[RGB]{0,0,255}{OSF Supplement}}) that visualizes the negative correlation between mileage and manufactured year,
five participants each identified different statistics---correlation, even distribution, N/A, negative correlation, and trend. We observed a similar case of four participants in a single-class line graph as shown in \autoref{fig:stimulus_sc}(b), where responses comprised positive correlation (2), compare (1), and trend (1). In \autoref{fig:stimulus_mc}(a), 
responses from five participants varied from the increase (1), upward-trend (1), similar trend (1), extremum (1), and compare-trend (1). Similarly, in \autoref{fig:teaser}(b), responses 
consisted of extremum (3), compare (3), count (2), and correlation (1).

 \subsubsection{Composition}
As discussed in Theme 1 (see Section \ref{sec-analysis_intention}), people's comprehension aligned best with designer intent using standard, single-class charts (44/72 complete or general match). With composited charts, alignment was highest with single-class juxtaposed (31/72) followed by multi-class non-juxtaposed (27/72) and finally, multi-class juxtaposed (24/72) graphs. One participant mentioned that these charts were significantly harder to use---``\textit{I could understand bar graphs and simple line graphs well but not scatterplots or graphs involving multiple categories or graphs.} [P16]" 

The majority of non-juxtaposed charts with single-class data effectively communicated their target properties (35/72). For example, participants noted a correlation in \autoref{fig:stimulus_sc}(b) as in ``\textit{There is a positive association between gross movie sales and IMDB. The association does not seem to be completely linear, but more dependent on the rating.} [P01],'' distribution and extremum in ``\textit{scatter plot showing max temp in Edoford during sunny days in a year, definite peak in summer from June to September or October} [P13]'' on \autoref{fig:stimulus_sc}(a), and correlation in ''\textit{Easier to read, shows the correlation between the turtle's body mass and flipper length per species in individual graphs} [P02]'' on \autoref{fig:stimuli_dimension}(a). 
These findings suggest that simple, single-class visualizations tended to better align with stated intentions; however, this alignment was still below 50\% overall. 

More complex data and conclusions drawn on them may require composite chart designs. Juxtaposition is a popular choice for composite charts, but its complexity led to more frequent misalignment and often a full lack of engagement with any specific patterns in the data. Single-class juxtaposed composition for both line (12/24) and bar (13/24) graphs aligned with their intended goals better than other compositions. For example, in \autoref{fig:stimuli_dimension}, participants identified the pattern and performed the comparative analysis of the data---more than 67\% of participants' responses aligned with designer's stated intention of comparing trends (5 times), comparing correlations (2 times) and estimating discrete values (1 time) on line and bar graphs respectively.

Viewers found single-class juxtaposed line graphs challenging when the graphs were vertically aligned. Eight responses mentioned that juxtaposed line graphs were ``hard'' to use, for example, in~\autoref{fig:stimulus_mc}.
One participant noted---``\textit{That final graph [single-class juxtaposed line graphs; sub-graphs are stacked on top of each other] was visually very hard for me to parse because all of the lines were stacked on top of one another. It was very hard for me to get a visual frame of reference for how to measure the growth of the stock.} [P07]''
This difficulty may have limited people's abilities to synthesize information across pairs of visualizations.  

Responses varied more between single- and multi-class versions of the same visualization type on juxtaposed graphs. For example, 45\% of responses matched designer intents in single-class juxtaposed line graphs versus 37\% in multiclass juxtaposed line graphs. We found an overall inverse correlation between the number of sub-graphs and alignment with the design intent. Specifically, fewer charts led to higher alignment. 
Non-juxtaposed charts tended to support higher alignment than other kinds of composite charts.

Multi-class juxtaposed graphs more often corresponded to comparisons with a less varied use of other statistics. 
Despite being data-rich, multiclass juxtaposed designs tended to lead to shallow interpretation---``\textit{In the left, I see plots showing the cost of a car and how old the car is associated with the odometer reading. On the right, I see three different colored graphs. I am assuming they are showing different car models.}" [P06] for \autoref{fig:stimulus_mc}(b).
Multi-class juxtaposed line graph responses focused more on local trends, such as \textit{change}~(7/24), \textit{increase-decrease}~(8), \textit{upward-downward trend}~(9), \textit{spike}~(2), \textit{stable trend}~(2), \textit{increase}~(10), and \textit{peak-valley}~(4). 

\subsubsection{Supplemental Graphical Information}
Several charts included basic annotations and added supplemental graphical elements. Although we did not control for the presence or absence of visual annotations (e.g., trend lines or highlights), instead choosing to mimic the widely-used graphs on 42 chosen datsets as closely as possible for ecological validity, we found that annotations assisted people in identifying the intended patterns in the data. Such annotations (e.g., highlighting value on a line graph, a visible baseline or grid line in bar graphs) helped people identify complex intentions, such as quarterly sales in a line graph and 52-week low stock price (see Fig. 29 in \href{https://osf.io/869ev/}{\textcolor[RGB]{0,0,255}{OSF Supplement}}) or the maximum budget beyond a threshold in \autoref{fig:stimulus_sc}. 
 Statistical baselines helped people identify the trend and correlations in all three types of graphs. However, in a line graph plotting stock data with \textit{52-week low} value as red baseline missed its intended task---highlight days in red when the stock hit below the 52-week low price and assess stock performance over time (correlation, trend, outliers, maxima, discrete value). Only one participant's response matched the intended tasks, reporting correlation and trend. Additionally, grid lines in bar graphs and line graphs helped people identify ranges and extract maximum values. 

 However, annotations also represent trade-offs. For example, gridlines can lead to attraction and repulsion effects that cause people to incorrectly estimate effect size \cite{xiong2019biased}. Further, calling out key information in a chart may cause people to fail to attend to other salient properties of the data \cite{xiong2019curse}. Future work should investigate the role annotation plays in shaping people's comprehension in the wild. 
%


\subsubsection{Individual Differences}
\label{sec-perparticipant}

Participant responses greatly varied between participants. 
As shown in \autoref{tab:resultcoding} (and Table 5 in Appendix B), 78\% (92/117) of complete match responses were grouped among only ten participants. 
59\% of participants did not provide responses that reliably aligned with the 
stated intentions. 
Only 45\% (11/24) of the participants managed to get an average (complete+general $>$ 7/12) to higher (complete+general $>$ 10/12) comprehension match with the 
stated intention. Three participants (Nursing, Environmental Science, and Computer Science major students) 
had 11/12 complete match responses, while three other participants had no responses that aligned with the stated intention at all.
The fact that only 34\% (8/24) participants had an upper high-range comprehension match shows that the designer's intended data communication was not immediately obvious to most readers despite these graphs replicating visualizations intended for mass communication.
This internal variance suggests that individual differences may play a role in what information a visualization naturally communicates. For example, people's performance differs with domain expertise \cite{hall2021professional} or graphical literacy~\cite{franconeri2021science}. While we did not systematically sample for specific demographic characteristics (e.g., profession, education, or graphical literacy), a comparative study of high-level visual comprehension on diverse populations requires further investigation.

\section{{Analysis of Design and Data Critiques}}
\label{sec-analysis_design_data}

Participants noted a range of additional factors that influenced their abilities to read a visualization. We identified and summarized these factors in \autoref{fig:design_data}. While not directly related to our core research questions, these observations provide additional insight into visualization design guidelines. 

 Several observations related to common graphic design principles, such as a visualization was ``\textit{easy to read because of [high] color contrast} [P16]" on \autoref{fig:teaser}(a); ``\textit{hard to read because of dashes} [P01]" or other less common visual metaphors on \autoref{fig:teaser}(c). 
Participants raised four common critiques based on a visualization's content: the need for \textit{additional information}, issues with \textit{legends and labels}, confusion with \textit{visual encodings}, and too much \textit{complexity}.

\noindent \textbf{Additional Information:} 
Participants frequently commented on missing information and context they felt limited their abilities to describe visualizations. 16 participants felt they needed additional information about the data to better comprehend a graph, such as units of measurement  (e.g., price in \autoref{fig:teaser}(c)),
or the demographics of people represented on graphs (Fig. 54 in supplements). 
Participants commented about acronyms used in the data, even when those acronyms were peripheral to patterns in the data. One participant noted, ``\textit{Some graphs are not easy to understand because I don't know what is IMDB} [P01]" in \autoref{fig:stimulus_sc}(b). People found  abbreviated stock names (\autoref{fig:teaser}(c)) or unfamiliar acronyms of airlines made charts more challenging to describe. Participants also found jargon off-putting, such as odometer (2/24), CPU architecture \& cores (5/24), and annuli (2/24).
One participant noted that ``\textit{Some of the graphs I had to concentrate on harder as I do not follow stocks or computer programming types.} [P15]'' 

\begin{figure}[t]
\centering
\includegraphics[width=0.8\columnwidth]{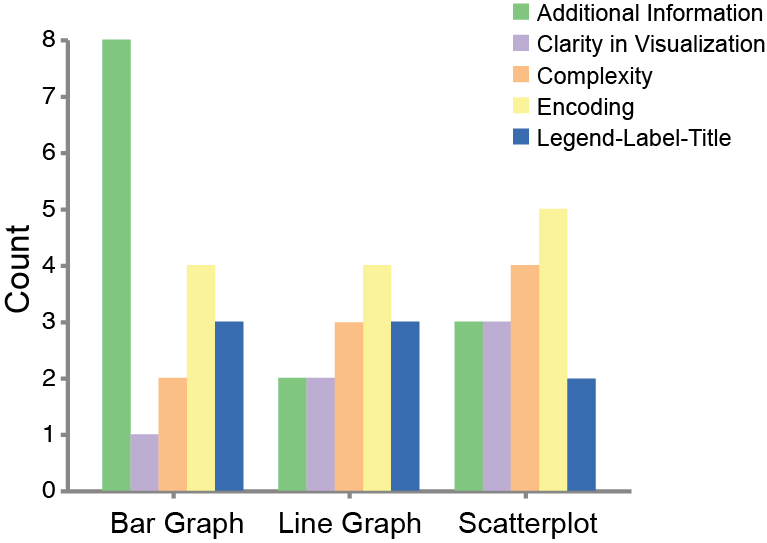}
\caption{Participants' responses and description of \textcolor[RGB]{157,81,160}{design critiques} and need for additional information on the \textcolor[RGB]{62,181,74}{data}. 
We observed that apart from additional information, all graph types received similar critiques.}
\label{fig:design_data}
\end{figure}

\noindent \textbf{Legends and Labels:} Participants actively attended to legends, labels, and titles, raising concerns when they were unclear (2/24) or missing (7/24). 
Having clear axis labels, titles, and legends is a known best practice in communicating quantitative information~\cite{tufte1985visual}. Our findings support these practices of including labels and legends considering the general audience, enabling them to identify patterns in the data.

\noindent \textbf{Visual Encoding:} Participants' comments about data encodings primarily centered on ``busy'' features of a graph, such as bright colors, dashed lines, and dots in graphs. People generally appreciated the use of color. For example, they noted that a graph may be ``easy to read because of color contrast.'' 

Color could also be misused, with one participant noting that visualization had---``\textit{Lots of colorful lines- it's a little jarring at first glance. There's a helpful key at the top though...} [P13]''
However, alternative encodings for delineating categories were not always as well-received. For example, all of the participants raised concerns about the dashed-line in \autoref{fig:teaser}(c), commenting that it was ``\textit{hard to read because of similar dash patterns} [P16]'' or ``\textit{Difficult to distinguish stroke dash, making the graph confusing.} [P01]'' Dashes are more robust across media and more accessible than colors; however, people found them more confusing. This contrast indicates a trade-off in categorical encoding that should be explored in future work. 

Colors that followed semantic guidance were also seen as helpful, in line with past recommendations~\cite{mukherjee2021context,lin2013selecting}. For example, in 
\autoref{fig:teaser}(b), where colors aligned with specific foods, participants' responses tended to align with the designer intentions: ''\textit{Output of specific foods (in tons) for specific food products, grouped by their food group. Almost all of the fruits had higher production than anything else, while dairy had the lowest productions.} [P18]''

\noindent \textbf{Clarity and Complexity:} 
Participants felt that graphs using too many distinct encodings were difficult to read. For example, \ref{fig:teaser}(a) was described as ``\textit{Hard to read because of the colorfulness and varying sizes of dots. Hard to estimate the insurance charges certain dot sizes correspond to...} [P13]'' 

Others noted that in some graphs ``\textit{too much information was given, making the graph hard to understand} [P21]'' or that there were
    ``\textit{too many elements in the graph being given, making it confusing} [P05]''

    Too much data using multiple visual channels increased perceived cognitive load, especially with juxtaposed graphs. 
    The think-aloud responses indicated that people perform graph comprehension in two steps, first processing legends, labels, mark encodings, and subgraph organization and then attending to patterns in the data. 
    Multiple encodings or sub-graphs complicated comprehension by making it harder to complete the first step, in line with our observations in Section \ref{sec-analysis_intention}.    
    However, thoughtful organization could increase perceived usability. ``\textit{The easier graphs to understand are the more organized and separated due to the clarity provided and lack of over stimulation to the brain; the messier and more colorful, the more my brain shuts down and doesn't want to process it,} [P21]."
    For example, Figure \ref{fig:teaser}(b) organizes like bars into the same physical group, encouraging people to consider groups as a single unit. A lack of organization led to a similar decrease in perceived usability. Participants found that ``\textit{dense clusters make it hard to read the given data points} [P14]" in scatterplots and that it was ``\textit{difficult to track the [target] line... due to all the lines overlapping and having similar colors} [P17]" in multi-class line graphs. 
     critiques from the participants on missing information and context about visualized data.

\section{Discussion}
\label{sec-discussion}

We investigated the high-level patterns people naturally see when encountering a visualization without a guiding task. This work provides preliminary steps towards characterizing the patterns people perceive unprompted from a graph as a function of its design. 
Our findings offer preliminary insight into the alignment of design intentions, reader intuitions, and guidelines from graphical perception and expert heuristics. Future research should explore 
these insights to create more refined heuristics that connect reader intuition with empirical findings.
\subsection{Key Themes}
Our thematic analysis offers a new lens for understanding visualization effectiveness by modeling the patterns and statistics people intuitively extract from a given visualization. 
This analysis revealed a misalignment between the guidelines established through experimental research on visualization and what individuals actually perceive when they encounter visualizations in the wild.
Our 
results 
highlight three key themes:

\vspace{3pt}\noindent \textcolor{orange}{\textbf{Intention does not align with comprehension.}}
The patterns that describe a visualization often failed to match its communication goals. The insights and salient patterns people intuitively extract from that visualization may not align with desired communication goals, even for visualizations reflecting best practices. In contrast to previous research, we noted that the degree of alignment is contingent on various factors, which we will discuss in 
subsequent subsections.

\vspace{3pt}\noindent  \textcolor{teal}{\textbf{Results from cued tasks may not alone predict the knowledge people build in a graph.}}
Visualization effectiveness is not fully captured by people's abilities to quickly and accurately complete a cued task as in traditional graphical perception paradigms. We need experiments emphasizing both top-down precision (i.e., with cued tasks) and bottom-up high-level visual comprehension (i.e., without cued tasks) to understand the knowledge people build in a graph.
 
\vspace{3pt}\noindent \textcolor{olive}{\textbf{Chart type alone is not sufficient to predict the information people extract from a graph. }}
The patterns people notice in a given visualization depend on several dimensions of visualization design. Mapping tasks to chart type is a powerful paradigm, but 
fails to account for all of the perceptual variables involved in visualization interpretation.

Our results offer a set of general considerations for visualization designers and researchers, 
helping reconcile prior findings (\autoref{sec-reflection}), 
providing implications for applying design heuristics (\autoref{sec-implication_Design}) and lending insight into methodological considerations for visualization evaluation (\ref{sec-implication_Methods}).

\subsection{\add{Relation to Prior Findings}}
\label{sec-reflection}

While specifically focused on statistical patterns perceived in the data, our findings complement observations from past work focusing on related elements of visualization use, such as preference and higher-level cognition and sensemaking~\cite{stokes2022striking, szafir2023visualization, burns2020evaluate}. 
\add{While these past findings address either specific audiences or complementary dependent measures, juxtaposing them against our results highlights two major considerations for visualization design that our findings confirm: the importance of multiple perspectives on ``effectiveness'' and the need to consider diverse audiences. }

\subsubsection{Defining Effectiveness}
The patterns people perceive in data are influenced by a range of factors, including the visual channels used~\cite{cleveland1984graphical,kim2018assessing}, the settings of those channels~\cite{kim2018assessing, szafir2018modeling}, the distribution and semantics of the data~\cite{song2018s,gramazio2014relation}, past knowledge about the data~\cite{xiong2019curse}, and even supplementary information like annotations and titles~\cite{kong2018frames}. Our results confirm that relying on cued statistical measures to assess performance provides important feedback but may not be sufficient to capture 
what a visualization actually communicates given this complex space.

Our findings resonate with previous work, highlighting the value of considering comprehension as a goal for interpreting communication intention~\cite{burns2020evaluate,north2006toward}
as cued tasks, particularly low-level tasks, alone may not suffice to capture the real communication goals and interpretations of visualizations~\cite{adar2020communicative,zacks1999bars}. 
These observations further confirm that encoding channels and graph types are insufficient to predict effectiveness~\cite{cheng2022captions,stokes2022striking}. Connecting findings across these different perspectives can help shape more holistic and robust guidance for effective visualization design, especially if paired with theoretical frameworks that offer grounded approaches to considering complex interactions between aspects of visualization design and use (e.g., rhetorical approaches~\cite{hullman2011visualization} or communication frameworks~\cite{adar2020communicative}). However, bringing these disparate sources of data and theory together under a unifying framework remains an open research challenge.

\subsubsection{Who is the Consumer?}
Our study aligns with Burns et al.'s call~\cite{burns2020evaluate} for evaluations that consider different levels of visualization understanding. The observed misalignment between visualization intentions and participant interpretation highlights the need for multi-tiered evaluation frameworks that assess both low-level task performance and high-level visual comprehension, which we discuss further in Section \ref{sec-implication_Methods}.
While our studies offer only a limited preliminary lens on comprehension, as our understanding of visualization comprehension grows, having these frameworks in place will allow the visualization community to act more readily upon the resulting findings. 

A key component of this analysis will be accounting for demographic factors that 
play a crucial role in how individuals interpret visualizations~\cite{peck2019data, wu2021understanding, carpendale2017subjectivity,grammel2010information}. Our findings confirm that comprehension varies significantly between individuals: to some degree, insight is in the eye of the beholder.
Even the concept of a 'novice' could be a multi-faceted label~\cite{burns2023we}, not only pertaining to one's unfamiliarity with visualization tools or techniques but also their ability to extract and interpret complex data patterns or even work with the same representations across data from different domains, as seen in hesitancy introduced by unfamiliar concepts or acronyms.
This variability may also inform improved literacy assessments and other methods for characterizing visualization comprehension. 
While not within the scope of this study, 
the effects of personal demographics and literacy on visual comprehension are critical to future work.

\subsection{Implications for Design}
\label{sec-implication_Design}
Effective designs make key patterns salient. Design practices use target tasks \cite{amar2005low,munzner2009nested} and problem domain information \cite{munzner2009nested} to drive a particular design choice. Our results indicate that this relationship may be powerful but insufficient: aspects of chart composition, data, and other design elements might alter the ``right'' visualization choice for a given task. For example, design guidelines suggest that \autoref{fig:teaser} and \autoref{fig:stimulus_mc} 
should communicate the target tasks, but participants did not reliably use those tasks to describe the content of the graphs.

Visualization designers should be aware that what they intend to communicate through a graph may not always align with what viewers naturally perceive. Relying solely on cued statistical measures to assess performance may not accurately determine whether their visualizations will successfully convey their objectives. This underscores the need for additional guidelines derived from future high-level comprehension studies to enhance the effectiveness of visualizations for a given communication goal.

Our study highlights that these design guidelines should go beyond simply mapping chart types to data types or statistics. Data complexity, composition, \add{and supplemental information such as captions~\cite{cheng2022captions}, titles~\cite{kong2018frames}, and additional text~\cite{stokes2022striking}} also influence how viewers interpret visualizations and what statistics they extract from those graphs. Designers should consider these factors when selecting visualization design.
Our results offer insights into trade-offs associated with common design choices, including: 

\noindent\textbf{Juxtaposition.} Past work provides conflicting perspectives on juxtaposition versus superposition~\cite{ondov2018face,javed2010graphical}. In our study, people found juxtaposed graphs more difficult to use, provided less synthesis of the data when describing the graphs, and were less likely to describe the graph's intended message. However, superimposed graphs (in our case, multiclass, non-juxtaposed) can also lead to increased visual complexity when communicating differences in classes with different visual channels, albeit less often than juxtaposed graphs in our results.
These conflicts suggest a need to understand better different types of complexity that may arise in 
more complex analysis scenarios.

\noindent \textbf{Expressiveness.} 
More complex graphs can be more expressive and communicate a broader range of information in a single chart~\cite{cleveland1986experiment} but were less likely to achieve their intended goals. People's descriptions of more complex graphs tended to focus on surface features and described fewer patterns and statistics, even though more questions could be answered with those charts. Designers should think about ways of encouraging engagement with more complex charts and help people more rapidly and effectively orient themselves within complex data. This finding also supports more minimalist design practice: Simplicity and clarity remain essential principles in visualization design.

\noindent \textbf{Chart Scaffolding.} Misinterpretations drawn from shallow, at-a-glance readings of misleading charts may suggest that people do not pay close attention to labels and legends~\cite{correll2017black}. However, past work in visual attention suggests that people prioritize axes and titles \cite{kim2017bubbleview}. We found that participants spent significant time understanding the visualization environment (e.g., graph itself, encodings, and text) and actively processed legends and labels.  Such labels should carefully consider their content as well: abbreviations, jargon, or missing units of measure  
can be distracting, alienate readers, and cause people to second guess their interpretations. 
    
\noindent\textbf{Supplemental Graphical Information.}
Additional visual cues to critical data or patterns, such as annotations, lead to more consistent interpretation. Designers can use these cues to direct attention to key aspects of the data and reduce ambiguity in data interpretation. Visualizations presenting data without statistical or written annotations, explicitly showing a target pattern, and missing contextual information (e.g., unit, topic, title, acronyms) led to greater variance in reported patterns. 
However, using explicit visual cues to highlight some patterns may cause people to miss other key takeaways \cite{xiong2019curse,boger2021jurassic}.

\subsection{Implications for Methods}
\label{sec-implication_Methods}
Conventional design guidelines in visualization often suggest which tasks a given visualization type is effective for, such as scatterplots for clustering, distribution characterization, and correlation. However, these guidelines are typically based on experiments where people are explicitly cued to identify specific statistics. 
Task effectiveness in controlled settings may not always translate to real-world scenarios, where people may 
approach visualizations without predefined goals in mind. Accompanying text and other chart contexts (e.g., labels and titles) may provide relevant cues in some cases~\cite{stokes2022striking,cheng2022captions,kong2018frames}. However, 
our results indicate a need for a broader range of methodological considerations in generalizing from empirical studies to design guidelines.

\subsubsection{Rethinking Graphical Perception}
Visualization effectiveness is typically measured through performance (e.g., accuracy, completion time, and error rate) and subjective experience (e.g., confidence, familiarity, and subjective preferences) in performing a specific task, such as estimating a statistic or finding an object~\cite{elliott2020design,quadri2021survey}. 
\textbf{Such cued experimental tasks may not adequately capture how people use visualizations in the wild} \add{or determine how well they achieve
their true communication goal}. \add{These results align with previous studies that suggest visual properties~\cite{stokes2022striking} or low-level tasks~\cite{adar2020communicative} alone are not sufficient to capture the designer's intention or determine the reader's interpretation.}
Visualizations are often contextualized in text with a descriptive title, caption, or accompanying prose that cues people to focus on the specific information. However, charts may also appear without this scaffolding, such as in presentations, exploratory analysis tools, or as ``teaser'' figures that people see before reading the accompanying article. 
Measuring performance using abstract statistical quantities may not tell designers whether visualizations in these contexts will achieve their goals. Further, failure to see a described pattern may lead to mistrust in information even when cued. 


A statistics-focused and directed research task may invoke a cognitive process known as the \textit{cognitivisation of perception} \cite{rogers2022cues}. This cognitivization may change how people engage with a visualization, causing them to note different patterns as being more salient than actually are~\cite{xiong2019curse}.
In practice, this implies that just because people can estimate a statistic from a given graph does not mean that they \textit{will} estimate that statistic, at least unprompted. Our findings offer preliminary steps towards designing for graphs in the wild without the risk of cognitivization. 

Future studies \add{adding visual comprehension evaluations} should complement traditional graphical perception approaches to explore methods for understanding how visualization design, data, and visual encodings 
can more universally predict whether a visualization will achieve its goals or \add{helped readers in interpreting the insights}. Doing so requires understanding both goal-directed (i.e., traditional graphical perception to understand accuracy and speed) and feature-driven (i.e., uncued comprehension to understand interpretation) visualization use.

\subsubsection{The Importance of Task Framing} 
\label{sec-taskframing}

Our pilot study only asked people to 
describe a visualization. Statistical tasks such as estimate correlation~\cite{harrison2014ranking}, identify outliers~\cite{albers2014task}, and characterize distribution~\cite{kim2018assessing} did not arise in these descriptions. As a result, we amended our study design to 
ask participants to identify questions each graph addressed. This new experimental task, in turn, led to significantly more statistical tasks and specific patterns emerging in people's responses. 
%
Best practices in survey design encourage asking about a target topic from different perspectives to avoid framing bias \cite{muller2015designing}. 
We found systematic differences in the ways people described the patterns in graphs depending on which experimental task (question or description) they were addressing.

\noindent \textbf{Question:} 
Asking people what questions one could answer with a graph led to more statistically-oriented responses, such as
``\textit{How do the sales of item X change throughout the year and the months?}'' or 
``\textit{How much, on average, does each genre need for production budget?}'' 
People frequently formed analytical questions with statistical quantities, 
and these responses tended to better align with the original goals of the graph: Question-oriented responses aligned in 182 of 288 responses, compared to 117 of 288 description-oriented responses
(\autoref{tab:resultcoding}).
\noindent \textbf{Description:} Asking people to describe a graph led to responses that were more deeply grounded in the graph's semantics, closer to traditional insights~\cite{north2006toward} \add{and level of understanding~\cite{burns2020evaluate}}. We found from the think-aloud that people initially spent time on understanding the graph environment (e.g., labels and legends) when describing the graph, leading to responses reflecting contextualized knowledge rather than raw statistics. 
However, for more complex graphs, descriptions are more often correlated with reading what one can physically see in the graph.
When describing a graph, people primarily browsed at a high level, and if they could not see salient patterns or statistics, they moved on. 
For example, 
one person verbally described a complex graph with large pauses between observations--``\textit{this graph has age of car and odometer reading on both graphs....[paused]...there difference price and age and model information....[paused]} [P05]" (see \autoref{fig:stimulus_mc}(b)). 

Prompting people to provide different kinds of information about a graph (i.e., descriptions versus questions) may allow experiments to tease apart the different levels of reasoning people engage within a graph. These tensions should be explored in future work to understand trade-offs for experimental evaluation.

\subsection{Limitations \& Future Directions}
\label{sec-limitation}
Our study emphasizes the need for further research to build a deeper understanding of high-level comprehension in data visualizations. This includes investigating new chart types, individual differences, data complexity, and the impact of design choices on data communication.

\paragraph{Design Factors:}
Our study evaluated the kinds of information people draw from scatterplots, line graphs, and bar graphs with different data types and compositions. While these reflect commonly studied visualization types \cite{quadri2021survey}, they are only a small portion of visualization types.
Future work should explore other visualization types, such as maps, pie charts, and bubble charts, as well as alternative designs for the current graphs. We synthesize our results across higher-level properties of a graph (e.g., its composition and visualization technique) to retain a manageable scope for our study; however, our results also indicate lower-level design factors like individual visual channels and their interactions also drive data interpretation. Future work should provide a more systematic investigation of how specific variables in visualization design affect high-level interpretation.
Our visualizations also draw from narrative visualizations to provide a ground truth for a graph's intended purpose, reflecting best practices. For future work, we plan to extend our method to exploratory scenarios, where knowledge evolves over time and, with use and visualizations, often aim to support a wider range of tasks.

\paragraph{Limitation Due to Variability in Results:}
We found a person's background, such as their education or profession, may affect their comprehension as 
in prior work \cite{hall2021professional}. Understanding the nature and extent of these differences can be instrumental in refining visualization design strategies to cater to diverse audiences effectively, \add{as demonstrated in past studies~\cite{grammel2010information,burns2023we}}. 
However, we did not systematically control for these factors across participants. \add{Moreover, our study did not aim to assess novices' performance in visual comprehension of widely-used visualizations. Hence, participant recruitment did not specifically target novices, contrary to suggestions in prior works~\cite{grammel2010information, burns2023we, peck2019data, burns2020evaluate}.}
While we believe our study reflects a typical audience for most applications, future work should be conducted on a larger population using systematic sampling methods to better understand how demographic factors influence data interpretation.

\paragraph{Study Procedure:} 
We focused on verbal and textual descriptions of graphs, but such descriptions omit additional data about what patterns are salient, such as what features in a graph people actively attend to. Future work can provide additional response 
data, such as eye-tracking, to further investigate the most salient elements of a graph and their influence on data interpretation.
We studied peoples' high-level comprehension of visualization via natural language and think-aloud protocols. However, previous studies~\cite{kong2019trust,borkin2015beyond} showed visual recognition and people's recall of the message in visualization also play a vital role in visual memory. Future work should focus on viewers' visual recall and examine what high-level comprehension people can recall from visualization and which designs may help people recall the messages. We believe these directions would help us further understand how to retain the key messages of a visualization.

\section{Conclusion}
\label{sec-conclusion}

Designers create visualizations to achieve specific high-level analytical or communication goals. These goals often require people to naturally extract complex, interconnected patterns in data. However, perceptual studies of visualization effectiveness focus on isolated, predefined, low-level tasks, such as estimating statistical quantities. 
People may efficiently assess that pattern when cued, but that pattern may not stand out when the same people encounter a similar visualization in the wild. 
We studied three visualization types---scatterplots, line graphs, and bar graphs---to investigate the high-level patterns people naturally see when they encounter a visualization. While graph type predominantly affected the statistics people noted, we found that interpretation varies with the data type, graph composition, and specific design elements.
These findings enable us to look at visualization design effectiveness \add{and their communication goals} with a new lens of high-level visual comprehension. \add{Through our results, we highlight the significance of incorporating both high-level comprehension and low-level tasks in assessing visualization effectiveness.}

\begin{acks}
This work was supported by the National Science Foundation under grant No. 2127309 to the Computing Research Association for the CIFellows project, NSF IIS-2046725, NSF IIS-1764089, and NSF IIS-2316496.
\end{acks}

\bibliographystyle{ACM-Reference-Format}
\bibliography{main}

\appendix
\section*{Appendix A: Pilot Study Information}
\label{appendix_A}

Appendix A introduced more details about our methodology and pilot study.

\subsection{Pilot Study}




\textbf{Task:} In this pilot study, we conducted an experiment to investigate what people intuitively see in data visualizations. We asked participants to describe what they saw in the given graphs. We recorded their verbal responses to evaluate whether the patterns participants saw matched the designers' goals as stated in the constituent articles.

\begin{figure*}[htbp]
    \centering
\includegraphics[width=0.9\linewidth]{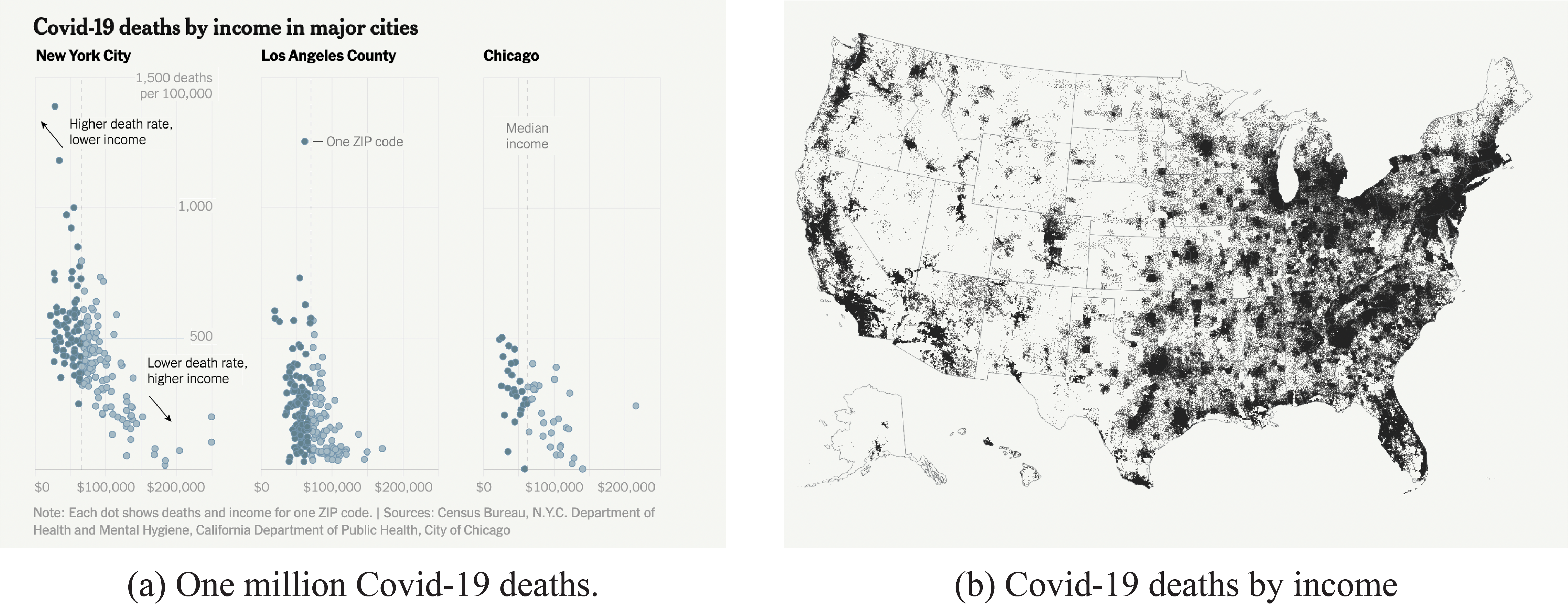}
\caption{Examples from New York Times. Graph (a) shows a USA map plotting one million Covid-19 deaths. Graph (b) demonstrates the COVID-19 deaths by income. }
\label{fig:casestudy}
\end{figure*}
\textbf{Study Setup:} We selected five different visualizations---one map, one area chart, two scatterplots (one being single class and the other multi-class), and one line graph, as listed in \autoref{tbl:3graph_summary}---from New York Times (see examples in \autoref{fig:casestudy}). We recruited 10 participants (six males, four females; 21-44 years of age) with varying levels of familiarity with visualizations, two of whom were academic research experts and the other eight more casual visualization users (five from academia and three from industry). We presented the visualizations in random sequential orders and asked the participants to \textit{"describe what you see in the graph"}. The entire study took no more than 10 minutes for each participant. We recorded their verbal responses and later evaluated whether the patterns that participants saw matched the designers' goals as stated in the constituent articles.

\begin{table}[htbp]
    \centering
    \caption{Information on the five graphs from New York Times used in the experiment. Links to articles and graphs are embedded in the description texts.}
    \label{tbl:3graph_summary}
    \begin{tabular}{c| c | p{2.5cm} | p{2.2cm} }
        ID & Visualization & Description & Visual Encoding  \\
        \hline
        \hline

        \quad  V1 & Map &  \href{https://www.nytimes.com/interactive/2022/05/13/us/covid-deaths-us-one-million.html}{Covid-19 1 million death} & Color saturation \\ %
         \quad  V2 & Area chart & \href{https://www.nytimes.com/interactive/2022/05/13/us/covid-deaths-us-one-million.html}{Covid-19 death per wave} & Area \\ 
         \quad  V3 & Scatterplot  & \href{https://www.nytimes.com/interactive/2022/05/13/us/covid-deaths-us-one-million.html}{Covid-19 deaths by income} & Position, Color \\ 
         
         \quad  V4 & Scatterplot & \href{https://www.nytimes.com/2020/01/02/learning/whats-going-on-in-this-graph-internet-privacy-policies.html}{Reading time for Popular text} & Position \\ 
        \quad  V5 & Line graph  & \href{https://www.nytimes.com/2019/02/28/learning/whats-going-on-in-this-graph-march-6-2019.html}{Use to tobacco products}  & Color, stroke or shading \\ 

    \end{tabular}

\end{table}

 \textbf{Result:} We discuss the results of two graphs from \autoref{fig:casestudy} categorizing participant responses as 1) understanding the graphs and 2) comments on the design, visualization, or analysis. Here we, are providing the detailed description only \textbf{map} (see \autoref{fig:casestudy} and \autoref{tbl:3graph_summary}).

 The majority of participants perceived density-based information from the map. However, participants did not identify the primary objective of the map: to convey that the COVID death crisis began in cities and spread to rural areas. Participants generally saw the COVID map as indicating regions of high and low density. Three participants (experts and a Ph.D. student) described the map as pointing to the highly-concentrated region (see \autoref{fig:casestudy} (a)), four noted a skew of data towards the east coast, and the remaining 3 participants instead noted that the east and the far west had a greater quantity or higher density of data. Four participants pointed out that the darker region represented higher quantities while the lighter region represented lesser quantities.

The participants who were experts talked in detail about how the design was misleading. For example, they mentioned that the graph \autoref{fig:casestudy} (a) lacked a scale to help them further interpret the approximate number of deaths in a given area. Five participants mentioned the lack of scale in the graph, and two participants asked for more information on the context of the graph and data. Six participants felt this graph missed conveying more information regarding quantity and scale. 

The designer's objective was to show that income is a predictor of Covid-19 mortality by showing a correlation between income and both death and vaccination rates in major cities. In line with this goal, six people described the graph as showing that Covid-19 deaths are more concentrated in lower-income regions of three cities (see \autoref{fig:casestudy} (a)), while one participant focused on correlations between death and income from NYC alone. Two participants compared the overall number of deaths across cities. Only two participants correctly interpreted data points as representing the death and income for a given zip code, with four participants instead seeing dot color as representing incomes as being either above or below the median.

For \autoref{fig:casestudy}, the responses varied among users when reflecting on the utility of the visualization. People felt the graphs were overloaded (three participants), complicated (two participants), the missing legend on colors (two participants), and were confused by the fact that each graph annotates different information (one participant).
 
\paragraph{Coding for the Formal Study:} Based on the participants' response and axial coding, we decided to code the formal study's response on-- \textit{comprehension match, statistical tasks, design critiques, and data critiques}. Additionally, we considered the graph composition of juxtaposition in the formal study. 

%


\label{sec-pilot}

\subsection{More Details of Formal Study}

\textbf{Low-level Task Definition:} Only those which are mentioned by participants.
\begin{enumerate}
    \item Correlation is defined as ``\textit{Given a set of data cases and two attributes, determine useful relationships between the values of those attributes.}'' The keywords from response used to code are: \textit {correlation, positive correlation, negative correlation, association, relation, not fully linear, prediction}.
    \item Trend is described as ``\textit{Given a set of temporal data cases, determine a pattern}.'' The keywords from response used to code are: \textit{increase, decrease, rate of change, change, drastic change, upward trend, downward trend, higher, lower}.
    \item Extremum is defined as ``\textit{Find data cases possessing an extreme value of an attribute over its range within the data set.}'' The keywords from response used to code are: \textit {maximum, minimum, peak, valley, high, low}.
    \item Compute Derive Value is described as ``\textit{Given a set of data cases, compute an aggregate numeric representation of those data cases.}'' The keywords from response used to code are: \textit{count, duration, mean}.
    \item Cluster is described as ``\textit{Given a set of data cases, find clusters of similar attribute values.}'' The keywords from response used in code are: \textit{cluster, grouped}.
    \item Characterize Distribution is described as ``\textit{Given a set of data cases and a quantitative attribute of interest, characterize the distribution of that attribute’s values over the set.}''  The keywords from response used to code are: \textit{even distribution, normal distribution, random, scattered data points, variability}.
    \item Anomalies is described as ``\textit{identify any anomalies within a given set of data cases concerning a given relationship or expectation, e.g., statistical outliers}'' The keywords from response used to code is: \textit {outlier}.
    \item Determine Range is described as ``\textit{Given a set of data cases and an attribute of interest, find the span of values within the set.}'' The keywords from response used to code are: \textit{from-to, over}.
    \item Compare is described as ``\textit{Given a set of data cases, compare any attributes within and between relations of the given set of data cases for a given relationship condition.}''  The keywords from the response used to code is: \textit {compare}.
\end{enumerate}

\section*{Appendix B: Metadata and Results}
\label{appendix_B}

%
%

Appendix B introduced the codebook, summarization of study results, the overall metadata, and several instances with different visual encodings.

\begin{table*}[ht]
\caption{Codebook example using a spreadsheet to code all users’ responses.}
\label{tab:codebook}
\scalebox{0.8}{
\begin{tabular}{|c|c|c|c|c|p{3cm}|p{3cm}|c|p{1.5cm}|c|c|c|}
\hline
PID & Name & Graph      & Data         & Composition    & Response-Description                                                & Response-Question                                            & Comprehension  & Statistics                            & Design & Data & Other \\
\hline
X   & IMDB  & Line graph & Single class & Non-juxtaposed & There is a positive association between gross movie sales and IMDB. The association does not seem to be completely linear, but is more dependent on the rating. & What's the association between gross movie sales and rating? Whether rating impact gross movie sales? & Complete Match & Association -\textgreater correlation & NA     & NA   & NA   \\
\hline
\end{tabular}
}
\end{table*}

\begin{table}[htbp]
\centering
  \caption{Summary of 24 participants' backgrounds and visualization experiences in the study. The top rows show participants' backgrounds, where 19 of 24 are students and 5 of 24 are working professions. The bottom rows are their reported visualization experiences.}
  \label{tab:participantback}
  \begin{tabular}{ccp{5.5cm}}
    \toprule
    \textbf{Fields} & \textbf{Count} & \textbf{Major / Profession }\\
    \midrule
    Humanities & 6/24 & Social Work, Law, European Studies, Journalism, Economics, Anthropology \\
    Computing & 6/24 & Computer Science (2), Health Informatics, Mathematics, Biomedical Engineering, Quantitative Biology \\
    Science & 3/24 & Psychology, Chemistry, Environmental Science \\
    Health & 4/24 & Nursing (2), Public Health, Pharmacology\\
    \midrule
    Professions & 5/24 & Health and Physical (2), Communication and Media Relations, Medical Technician, Educational Technology,  \\
    \toprule
    
    \textbf{Experience} & \textbf{Count} & \textbf{Details}\\
    \midrule
    Extensive & 1/24 & Almost an expert in visualization \\
    Casual & 8/24 & Usually use visualizations in projects \\
    Little & 13/24 & Created some visualizations before \\
    No & 2/24 & Almost didn't use visualizations before \\
  \bottomrule
\end{tabular}
\end{table}

\autoref{tab:participantback} shows details of all the participants' demographics in our study.
\autoref{tab:field_res} shows the summary results of participants' comprehension per participant's field of education (students) and working professionals.
\autoref{tab:perpartcipant} shows the summary results of participants' comprehension match per match level in the Description task.
\autoref{tab:abbr} illustrates the abbreviations used in the rest results of our analysis.
\autoref{tab:metatable} shows all of the metadata of our study.
\autoref{fig:count} shows the summary counts of questions participants responded to for the Question Task per graph type and data type.

\begin{figure}[htbp]
\centering
\includegraphics[width=0.9\columnwidth]{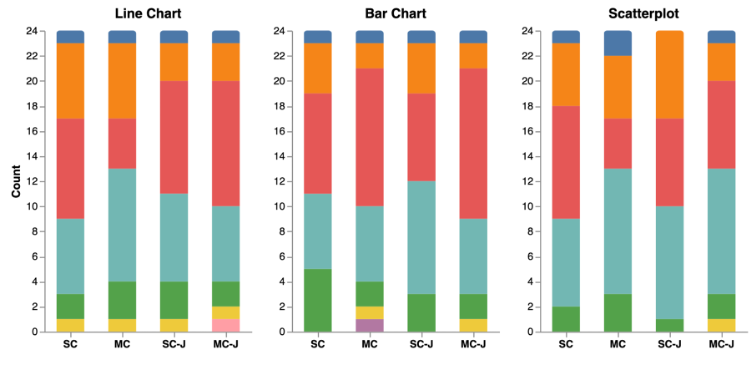}
\caption{Summary counts of questions participants responded to the Question Task.}
\label{fig:count}
\end{figure}

\begin{table}[htbp]
    \centering
    \caption{Participant comprehension summary results as per participant's field of education (students) and working professional.}
    \label{tab:field_res}
    \begin{tabular}{ |c|cccc|  }
     \hline
     Field of Study & CM & GM & PM & NM \\
     \hline
     Computing \& Related & \cellcolor{yellow!43}31  & \cellcolor{yellow!12}8 & \cellcolor{yellow!19}15 & \cellcolor{yellow!6}6 \\
     Science  & \cellcolor{yellow!25}31  & \cellcolor{yellow!1}2 & \cellcolor{yellow!3}7 & \cellcolor{yellow!8}8 \\
     Health & \cellcolor{yellow!13}17  & \cellcolor{yellow!1}3 & \cellcolor{yellow!7}10 & \cellcolor{yellow!18}18 \\
     Humanities  & \cellcolor{yellow!20}21  & \cellcolor{yellow!7}9 & \cellcolor{yellow!17}23 & \cellcolor{yellow!20}19 \\
     \hline
     Working Professional  & \cellcolor{yellow!17}17  & \cellcolor{yellow!4}4 & \cellcolor{yellow!11}12 & \cellcolor{yellow!28}27 \\
     \hline
     Count (Total is \textbf{288})  & \cellcolor{gray!90}117  & \cellcolor{gray!25}26 & \cellcolor{gray!53}67 & \cellcolor{gray!60}78 \\
     \hline  
    \end{tabular}
\end{table}

\begin{figure}[htbp]
\centering
\includegraphics[trim=0 8pt 0 0, clip, width=1\columnwidth]{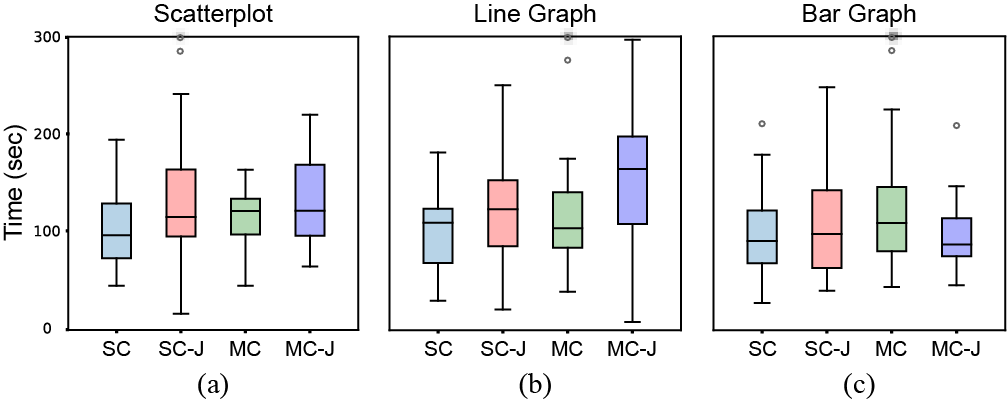}

\hspace{0.07\columnwidth}
\subfloat[\label{fig:time:a}]{\hspace{0.3\columnwidth}}
\subfloat[\label{fig:time:b}]{\hspace{0.3\columnwidth}}
\subfloat[\label{fig:time:c}]{\hspace{0.3\columnwidth}}

\caption{Average response time for four visualization types. SC: single-class; MC: multi-class; SC-J: single-class juxtaposed; MC-J: multi-class juxtaposed. The dark transparent shadows mean this value exceeds the upper bound (300s), and a larger range of shadows means more outliers. As we can see participants took more time to comprehend multi-class graphs in all three graphs and multi-class-juxtaposed in scatterplot and line graphs. The maximum average comprehension time is for multi-class-juxtaposed line graphs.}
\label{fig:time}
\end{figure}

\begin{table}[htbp]
\centering
\caption{Participants' comprehension match summary by categories on the Description task (see Sect. 3.1 in the paper). Each participant saw 12 different graphs in the study. Ranked by the number of Complete Match. }
\label{tab:perpartcipant}
\begin{tabular}{|c|cccc|c|}
\hline  
Assigned & \multicolumn{4}{c|}{Comprehension} & \multirow{2}{*}{Total} \\
\cline{2-5}
ID & CM & GM & PM & NM & \\
\hline  
05 & 11 & 1 & 0 & 0 & 12\\
19 & 11 & 0 & 1 & 0 & 12\\
22 & 11 & 0 & 1 & 0 & 12\\
09 & 10 & 1 & 1 & 0 & 12\\
10 & 10 & 0 & 2 & 0 & 12\\
20 & 9 & 1 & 2 & 0 & 12\\
08 & 9 & 1 & 1 & 1 & 12\\
17 & 8 & 2 & 1 & 1 & 12\\
12 & 7 & 2 & 1 & 2 & 12\\
01 & 6 & 1 & 4 & 1 & 12\\
13 & 6 & 1 & 2 & 3 & 12\\
18 & 4 & 2 & 3 & 3 & 12\\
16 & 4 & 1 & 3 & 4 & 12\\
07 & 3 & 2 & 3 & 4 & 12\\
11 & 3 & 0 & 2 & 7 & 12\\
06 & 2 & 1 & 4 & 5 & 12\\
21 & 2 & 0 & 4 & 6 & 12\\
14 & 1 & 2 & 6 & 3 & 12\\
24 & 0 & 4 & 3 & 5 & 12\\
02 & 0 & 3 & 9 & 0 & 12\\
03 & 0 & 1 & 7 & 4 & 12\\
04 & 0 & 0 & 4 & 8 & 12\\
23 & 0 & 0 & 2 & 10 & 12\\
15 & 0 & 0 & 1 & 11 & 12\\
\hline
\cellcolor{gray!24}Total & \cellcolor{gray!24}117 & \cellcolor{gray!24}26 & \cellcolor{gray!24}67 & \cellcolor{gray!24}78 & \cellcolor{gray!24}288 \\ 
\hline
\end{tabular}
\end{table}


\begin{table*}
\centering
\vspace{-1em}
\scalebox{0.6}{
\begin{tabular}{c|cccc|cccc|cccccccccc|ccccccccc}
  & \multicolumn{4}{c|}{Class Type} & \multicolumn{4}{c|}{Comprehension Match} & \multicolumn{10}{c|}{Visual Encoding} & \multicolumn{9}{c}{Statistics} \\
\hline
Abbr. & SC & SC-J & MC & MC-J & CM  & GM  & PM  & NM & C  & D  & P & S  & T & W  & BH & BL & CS & SD & Corr & Comp & Tr  & Clu  & CDV & DR  & Xtrm  & CD  & Anom     \\
\hline
Name  & \rotatebox{270}{Single-Class} & \rotatebox{270}{Single-Class Juxtapose} & \rotatebox{270}{Multi-Class} & \rotatebox{270}{Multi-Class Juxtapose} & \rotatebox{270}{Complete Match} & \rotatebox{270}{General Match} & \rotatebox{270}{Partial Match} & \rotatebox{270}{No Match} & \rotatebox{270}{Color} & \rotatebox{270}{Dots} & \rotatebox{270}{Position} & \rotatebox{270}{Size} & \rotatebox{270}{Texture} & \rotatebox{270}{Text} & \rotatebox{270}{Bar Height} & \rotatebox{270}{Bar Length} & \rotatebox{270}{Color + Semantics} & \rotatebox{270}{Stroke Dash} & \rotatebox{270}{Correlation} & \rotatebox{270}{Comparison} & \rotatebox{270}{Trend} & \rotatebox{270}{Cluster} & \rotatebox{270}{Compute Derived Value} & \rotatebox{270}{Determine Range} & \rotatebox{270}{Extremum} & \rotatebox{270}{Characterize Distribution} & \rotatebox{270}{Anomaly} \\
\end{tabular}
}
\caption{Summary of abbreviations used in \autoref{tab:metatable}.}
\label{tab:abbr}
\vspace{-2em}
\end{table*}

\begin{table*}[htbp]
\centering
\scalebox{0.75}{%
\begin{tabular}{|c|c|c|c|c|c|cccc|cc|c|}
\hline
\rowcolor[rgb]{0.753, 0.753, 0.753}{\cellcolor[rgb]{0.753, 0.753, 0.753}} & {\cellcolor[rgb]{0.753, 0.753, 0.753}} & {\cellcolor[rgb]{0.753, 0.753, 0.753}} & {\cellcolor[rgb]{0.753, 0.753, 0.753}} & {\cellcolor[rgb]{0.753, 0.753, 0.753}} & {\cellcolor[rgb]{0.753, 0.753, 0.753}} & \multicolumn{4}{c}{\textbf{Comprehension}} & \multicolumn{2}{c}{\textbf{Critique}} & {\cellcolor[rgb]{0.753, 0.753, 0.753}} \\
\rowcolor[rgb]{0.753, 0.753, 0.753}\multirow{-2}{*}{{\cellcolor[rgb]{0.753, 0.753, 0.753}}\textbf{Graph-type}} & \multirow{-2}{*}{{\cellcolor[rgb]{0.753, 0.753, 0.753}}\textbf{Data-type}} & \multirow{-2}{*}{{\cellcolor[rgb]{0.753, 0.753, 0.753}}\textbf{ID}} & \multirow{-2}{*}{{\cellcolor[rgb]{0.753, 0.753, 0.753}}\textbf{Dataset}} & \multirow{-2}{*}{{\cellcolor[rgb]{0.753, 0.753, 0.753}}\textbf{Encoding}} & \multirow{-2}{*}{{\cellcolor[rgb]{0.753, 0.753, 0.753}}\textbf{Count}} & {\cellcolor[rgb]{0.682, 0.867, 0.506}}CM & {\cellcolor[rgb]{0.682, 0.867, 0.506}}GM & {\cellcolor[rgb]{0.682, 0.867, 0.506}}PM & {\cellcolor[rgb]{0.973, 0.91, 0.537}}NM & \multicolumn{1}{l}{{\cellcolor[rgb]{0.133, 0.882, 0.922}}\textbf{\textbf{Dataset}}} & \multicolumn{1}{l}{{\cellcolor[rgb]{0.133, 0.882, 0.922}}\textbf{\textbf{Design}}} & \multirow{-2}{*}{{\cellcolor[rgb]{0.753, 0.753, 0.753}}\textbf{Statistics}} \\
\hline
\multirow{20}{*}{\rotatebox{90}{Scatterplot}} & \multirow{5}{*}{SC} & 1 & \emph{Boeing} & P & 5 & 3 & 0 & 1 & 1 & 3 & 0 & Corr (3), CD (2), Tr (1) \\
 &  & 2 & $CO_2$ & P & 6 & 2 & 1 & 1 & 2 & 0 & 2 & Corr (2), Clu (1), Comp (1) \\
 &  & 3 & \emph{Medical} & P & 4 & 1 & 0 & 2 & 1 & 0 & 0 & Xtrm (2), CDV (1)  \\
 &  & 4 & \cellcolor{orange}\emph{Sunny day} & P & 4 & 2 & 2 & 0 & 0 & 1 & 0 &  Tr (2), Xtrm (1), CD (1) \\
 &  & 5 & \emph{Turtle} & P & 5 & 4 & 0 & 0 & 1 & 4 & 0 & Corr (4), Clu (1), DR (1), CDV (1) \\
\cline{2-13}
 & \multirow{5}{*}{MC} & 6 & \emph{Activity} & C, P & 6 & 1 & 2 & 2 & 1 & 2 & 1 & Xtrm (1), CDV (1), Clu (2), Comp (1) \\
 &  & 7 & \emph{Car models} & C, S, P & 7 & 4 & 0 & 1 & 2 & 1 & 0 & Corr (6), Clu (2), DR (1), Xtrm (2) \\
 &  & 8 & \emph{Horsepower} & C, P & 4 & 2 & 1 & 0 & 1 & 1 & 2 & Corr (4), Tr (1) \\
 &  & 9 & \emph{TempChange} & C, P & 4 & 1 & 0 & 3 & 0 & 1 & 1 & Tr (5) \\
 &  & 10 & \emph{Penguin} & C, P & 3 & 3 & 0 & 0 & 0 & 2 & 0 & Corr (2), Comp (1) \\
\cline{2-13}
 & \multirow{5}{*}{SC-J} & 11 & \emph{CPU} & C, P & 6 & 2 & 0 & 2 & 2 & 2 & 0 & Corr (4), DR (1), Tr (1) \\
 &  & 12 & \emph{Horsepower} & C, P & 5 & 1 & 1 & 2 & 1 & 2 & 0 & Corr (4), Comp (1), CD (1) \\
 &  & 13 & \emph{Penguin} & C, P & 5 & 1 & 0 & 2 & 2 & 1 & 1 & Corr(3), Comp(1) \\
 &  & 14 & \emph{PM 2.5} & C, P & 2 & 1 & 0 & 1 & 0 & 2 & 0 & Clu (1), CD (1), Comp (1) \\
 &  & 15 & \emph{Tumor} & C, P & 6 & 3 & 0 & 1 & 2 & 5 & 0 & Comp (3), Clu (1), CD (2), Corr (1) \\
\cline{2-13}
 & \multirow{5}{*}{MC-J} & 16 & \emph{Age-BMI} & C, P & 4 & 1 & 0 & 0 & 3 & 1 & 2 & CD (1), Corr (1) \\
 &  & 17 & \cellcolor{magenta}\emph{Car models} & C, P, S & 5 & 0 & 0 & 1 & 4 & 0 & 0 & Comp (1) \\
 &  & 18 & \emph{Weather} & CS, P & 6 & 2 & 1 & 1 & 2 & 4 & 1 & Tr (6), Xtrm (1) \\
 &  & 19 & \cellcolor{lime}\emph{Insurance} & C, P, S & 3 & 1 & 0 & 2 & 0 & 2 & 1 & Corr (2), Comp (1) \\
 &  & 20 & \emph{Titanic} & C, P & 6 & 4 & 1 & 1 & 0 & 3 & 1 & Comp (5), Corr (1) \\
\hline
\multirow{20}{*}{\rotatebox{90}{Line graph}} & \multirow{5}{*}{SC} & 21 & \emph{Google} & P, D & 6 & 4 & 1 & 0 & 1 & 3 & 0 & Tr (6) \\
 &  & 22 & \emph{Gov bond} & P & 5 & 2 & 1 & 2 & 0 & 3 & 0 & Tr (5), Xtrm (3) \\
 &  & 23 & \cellcolor{orange}\emph{IMDB} & P & 3 & 2 & 0 & 1 & 0 & 1 & 0 & Corr (4), Tr (2) \\
 &  & 24 & \emph{Sales} & P & 5 & 3 & 1 & 1 & 0 & 5 & 0 & Xtrm (1), Tr (6), CDV (1), Comp (1) \\
 &  & 25 & \emph{Stock red} & C, P, SD & 5 & 1 & 0 & 0 & 4 & 1 & 0 & Tr (3) \\
\cline{2-13}
 & \multirow{5}{*}{MC} & 26 & \emph{5-gov} & C, P & 6 & 4 & 0 & 1 & 1 & 4 & 2 & Tr (5), Comp (1), Xtrm (1) \\
 &  & 27 & \emph{EEG} & C, P & 2 & 0 & 0 & 0 & 2 & 0 & 0 & Xtrm (2), Tr (1) \\
 &  & 28 & \emph{Titles} & C, W, P & 7 & 1 & 0 & 4 & 2 & 2 & 1 & Tr (2), CDV (2), Xtrm (1), Corr (1) \\
 &  & 29 & \emph{Stock red} & C, SD, P & 4 & 1 & 0 & 1 & 2 & 0 & 2 & Corr (1), Tr (1) \\
 &  & 30 & \cellcolor{lime}\emph{Stock} & SD, P & 5 & 2 & 0 & 1 & 2 & 1 & 1 & Tr (4), Comp (1) \\
\cline{2-13}
 & \multirow{5}{*}{SC-J} & 31 & \emph{Airline} & C, P & 6 & 2 & 1 & 1 & 2 & 2 & 0 & Tr (4) \\
 &  & 32 & \emph{IMDB} & C, P & 4 & 4 & 0 & 0 & 0 & 2 & 0 & Corr (3), Comp (1), Tr (1) \\
 &  & 33 & \emph{Visitors} & P & 4 & 2 & 0 & 0 & 2 & 2 & 0 & CDV (1), Comp (2) \\
 &  & 34 & \emph{Spotify} & C, P & 5 & 2 & 0 & 1 & 2 & 1 & 1 & CDV (2), Tr (3) \\
 &  & 35 & \emph{Stock} & C, P & 5 & 1 & 0 & 1 & 3 & 1 & 1 & Xtrm (2), Tr (1), \\
\cline{2-13}
 & \multirow{5}{*}{MC-J} & 36 & \emph{Activity-covid} & C, P & 6 & 3 & 1 & 2 & 0 & 0 & 0 & Tr (5), Comp (3), CDV (1) \\
 &  & 37 & \emph{Unemployed} & C, P & 3 & 2 & 0 & 0 & 1 & 2 & 0 & Tr (3), Comp (1) \\
 &  & 38 & \emph{Covid-3} & C, P & 5 & 1 & 0 & 1 & 3 & 0 & 3 & Comp (2), CDV (1), Tr (1) \\
 &  & 39 & \cellcolor{magenta}\emph{Income} & C, P & 8 & 1 & 0 & 5 & 2 & 4 & 0 & Tr (8), Xtrm (2), Comp (1) \\
 &  & 40 & \emph{Working} & C, P & 2 & 2 & 0 & 0 & 0 & 2 & 0 & Tr (1), Comp (1) \\
\hline
\multirow{20}{*}{\rotatebox{90}{Bar graph}} & \multirow{5}{*}{SC} & 41 & \emph{12-cat products} & CS, BH, P & 6 & 2 & 1 & 1 & 2 & 1 & 1 & Xtrm (2), Comp (2), CDV (1)  \\
 &  & 42 & \emph{Afghan} & BH, P & 5 & 2 & 0 & 2 & 1 & 0 & 1 & Tr (4)  \\
 &  & 43 & \emph{TempChange} & C, P & 5 & 3 & 1 & 0 & 1 & 3 & 1 & Tr (6)  \\
 &  & 44 & \cellcolor{orange}\emph{Budget} & BL, P & 5 & 3 & 1 & 1 & 0 & 5 & 1 & Xtrm (8), Comp (2)  \\
 &  & 45 & \emph{Electricity} & BH, P & 3 & 1 & 0 & 0 & 2 & 1 & 0 & Comp (1)  \\
\cline{2-13}
 & \multirow{5}{*}{MC} & 46 & \emph{Movie} & C, P, W & 4 & 1 & 2 & 0 & 1 & 2 & 0 & Comp (2)  \\
 &  & 47 & \emph{ProgProficiency} & C, P, BH & 5 & 3 & 0 & 2 & 0 & 1 & 0 & Corr (2), CDV (1), Comp (2)  \\
 &  & 48 & \emph{Textured-Alt}  & T, P, BL & 5 & 0 & 1 & 2 & 2 & 2 & 0 & CDV (1), DR (1), Comp (3)  \\
 &  & 49 & \emph{Textured} & T, P, BH & 5 & 0 & 2 & 3 & 0 & 0 & 2 & Xtrm (1), CDV (2), Comp (1)  \\
 &  & 50 & \emph{Visitors} & C, P, BH & 5 & 4 & 0 & 0 & 1 & 4 & 0 & Xtrm (6), Anom (3), Comp (1) \\
\cline{2-13}
 & \multirow{5}{*}{SC-J} & 51 & \emph{Revenue} & C, P, BH & 3 & 2 & 0 & 1 & 0 & 1 & 1 & Xtrm (2), Corr (3), Comp (1)  \\
 &  & 52 & \emph{Success} & C, P, BL & 5 & 1 & 0 & 1 & 3 & 0 & 2 & Xtrm (2), Corr (1), Comp (1)  \\
 &  & 53 & \emph{Avg sales} & C, P, BH & 5 & 3 & 0 & 1 & 1 & 2 & 1 & Xtrm (6), Comp (1), Tr (1)  \\
 &  & 54 & \emph{Media} & C, P, BH & 5 & 3 & 1 & 1 & 0 & 2 & 1 & Comp (3), CDV (1), Xtrm (3), CD (1)  \\
 &  & 55 & \emph{Twitter} & C, P, BH & 6 & 3 & 0 & 2 & 1 & 3 & 0 & Comp (5)  \\
\cline{2-13}
 & \multirow{5}{*}{MC-J} & 56 & \cellcolor{lime}\emph{Products} & C, P, BH & 5 & 3 & 0 & 0 & 2 & 1 & 0 & Xtrm (5), Comp (1)  \\
 &  & 57 & \emph{Farms} & C, P, BL & 7 & 3 & 1 & 1 & 2 & 2 & 1 & Xtrm (2), Comp (1), CDV (1),  \\
 &  & 58 & \emph{ProgPopularity} & C, P, BH & 2 & 0 & 0 & 1 & 1 & 1 & 0 & Xtrm (2), Comp (1)  \\
 &  & 59 & \emph{Q\&A} & C, P, BL & 3 & 1 & 2 & 0 & 0 & 1 & 0 & Xtrm (2), Comp (1)  \\
 &  & 60 & \emph{Stock} & C, P, BH & 7 & 0 & 0 & 3 & 4 & 2 & 2 & Xtrm (2), Tr (2), Comp (2) \\ \hline
\end{tabular}
}
\caption{Summary of the visualizations shown to users in our user study, grouped by 3 graph types and 4 class types and thus forming 12 categories, each containing 5 visualizations. Check \autoref{tab:abbr} for definitions of abbreviations. {\color{lime}Lime} datasets are instances shown in \autoref{fig:teaser} in the paper, {\color{orange}orange} datasets are in \autoref{fig:stimulus_sc}, and {\color{magenta}magenta} datasets are in \autoref{fig:stimulus_mc}.}
\vspace{-2em}
\label{tab:metatable}
\end{table*}

\clearpage

\onecolumn
\begin{longtable}{|c|c|c|p{12.8cm}|}
\caption{Summary of the references and stated objectives of each stimulus.}
\label{tab:source}\\
\hline
ID & Dataset & Reference & Stated Objective     \\ \hline
1  & \emph{Boeing}          &    \cite{boeing}    & To show and demonstrate the correlation between the mileage and year of production of a set of Boeing airplanes.   \\ \hline
2  & $CO_2$                                 & \cite{owidco2andgreenhousegasemissions}    & To show a set of countries’ per capita GDP and their annual CO2 emission and demonstrate the quantity of CO2 emissions concentrated in countries with low per capita income.  \\ \hline
3  & \emph{Medical}         &   \cite{cms2022health}     & To show the distribution of age and medical expenditure of a set of people to understand how one relates to another (correlation).      \\ \hline
4  & \emph{Sunny day}       &   \cite{sunnydays}     & To document the pattern in the days where sunny weather occurs and the maximum temperature of each recorded day in a city from January to May and from August to December in 2012.  \\ \hline
5  & \emph{Turtle}          &  \cite{turtles} & To document turtles’ mass and their numbers of annuli and demonstrate the correlation between the two.                                       \\ \hline
6  & \emph{Activity}        &   \cite{activities}   & To plot the distribution of the number of calories burned vs. time from three workout activities. Also, show the clusters formed by the distribution pattern of burnt calories vs. time in 3 different colors.                \\ \hline
7  & \emph{Car models}      &   \cite{carmodels}     & To show the correlation between car odometer readings and their ages. Additionally, with two categorical encodings (color and point size) compare the car's odometer reading and age with prices, and models. Demonstrate the correlation between the factors and also see how one predicts the car price for a given age based on the odometer reading. The distribution or relation varies between different models.                                        \\ \hline
8  & \emph{Horsepower}      &   \cite{MPG_2017}     & To show the correlation between miles per gallon and horsepower of cars and also compared the distribution and correlation of these two variables for cars manufactured in 3 different countries.    \\ \hline
9  & \emph{TempChange}      &     \cite{tempchange}   & To show and compare the changes in US temperature over roughly one and a half centuries using dots whose positions and colors correspond to the change in temperature it indicates. It easily indicates how temperature has risen in one and half centuries.    \\ \hline
10 & \emph{Penguin}         &   \cite{horst2020palmerpenguins}     & To show the correlation between body mass and flipper length of 3 species of penguins. It shows a positive correlation between mass and flipper length. Penguin species can be identified by physical appearance (mass length) and each of their distribution varies.           \\ \hline
11 & \emph{CPU}             &     \cite{cpudata}   & To show the correlation between time vs no of CPU/core and compare between 6 distinct architectures with six juxtaposed graphs. We showed the completion time for the same task on different numbers of cores, with each graph showing only CPUs of the same architecture.                       \\ \hline
12 & \emph{Horsepower}      &    \cite{MPG_2017}    & To show and compare the miles per gallon and horsepower of cars manufactured in 3 different countries with three different Scatterplots, with each graph showing cars manufactured in different countries. We want to represent the correlation and distribution between MPG and horsepower for the given car.                                         \\ \hline
13 & \emph{Penguin}         &   \cite{horst2020palmerpenguins}     & To show the correlation between body mass and flipper length of 3 species of penguins. It shows a positive correlation between mass and flipper length. Penguin species can be identified by physical appearance (mass length) and each of their distribution varies.          \\ \hline
14 & \emph{PM 2.5}          &   \cite{pm25}     & To show and compare hundreds of PM 2.5 density readings recorded on two days in 3 different cities, with each separate graph showing readings from the different cities. Understand individual city distribution and compare three.                                                \\ \hline
15 & \emph{Tumor}           &   \cite{ghazal2022detection}   & To show and compare the radius and concavity of benign and malignant cancer tumors in two separate scatterplots, with each graph showing either benign tumors or other malignant ones.          \\ \hline
16 & \emph{Age-BMI}         &   \cite{bmiinsurance}     & To show and compare the age and BMI of a set of clients from 4 different regions show the distribution between patient's age and their BMI and compare the same for the information in four different regions in separate scatterplots. We also demonstrated the gender of insurers with different colors.   We want to demonstrate the distribution between these factors and compare them among 4 regions. \\ \hline
17 & \emph{Car models}      &    \cite{carmodels}    & To show the correlation between car odometer readings, and car ages, in two side-by-side scatterplots. Compare the two graphs on the price of the car, and models of a set of cars, with one graph showing cars’ prices and the other showing their models.          \\ \hline
18 & \emph{Weather}         &   \cite{weather}     & To document and demonstrate the trend/pattern in days for 5 types of weather occur and the maximum temperature of each recorded day in a city from 1) January to May (one graph) and from 2) August to December in 2012   (another graph). We want to show the trend in temperature for the two seasons.                                     \\ \hline
19 & \emph{Insurance}       &    \cite{bmiinsurance}    & To show the distribution between patients' age and their BMI and compare the same for the information in four different regions in separate scatterplots. We separately encoded the gender of the insurer and their insurance premium amount. We want to demonstrate the distribution between these four factors and compare them among 4 regions.                                                  \\ \hline
20 & \emph{Titanic}         &    \cite{titanic}    & To show and compare the distribution of survival/death vs. fare of Titanic passengers separated into 2 sets with one consisting only of males and the other only of females.         \\ \hline
21 & \emph{Google}          &   \cite{stock}     & To document the stock price of Google roughly from 2004 to 2010 in order to show patterns in stock price changes.      \\ \hline
22 & \emph{Gov bond}        &    \cite{govbond}    & To document the pattern in the yields of long-term government bonds over roughly 6 decades.                                            \\ \hline
23 & \emph{IMDB}            &   \cite{imdb}     & To document the correlation between movies’ IMDB ratings and their revenues made in the   US.                                            \\ \hline
24 & \emph{Sales}           &    \cite{protainsales}    & To document the pattern of the sales figures of a protein product from 2006 to 2007.                                                     \\ \hline
25 & \emph{Stock red}       &    \cite{stock}    & To document the trend (high and low) in the price of a stock over May 2021 and highlight in red days on which its price hit below a threshold.      \\ \hline
26 & \emph{5-gov}           &   \cite{govbond}     & To show the pattern and distribution of annual performance of government bonds issued by 5 countries and highlight years in which the performance hit below 0\% return.                                   \\ \hline
27 & \emph{EEG}             &   \cite{rosen2020linesmooth}   & To show the pattern of EEG Readings of 500 samples on 3 channels and compare their similarities by superimposition.                     \\ \hline
28 & \emph{Titles}          &   \cite{titles}     & To show and compare the number of grand slam titles owned by 5 tennis players at different ages.                        \\ \hline
29 & \emph{Stock red}       &  \cite{stock}      & To show the trend in prices of 4 stocks in May 2021 and highlight days on which a   certain stock’s price hit below a threshold of 52-week-low. The price below the threshold is shown in red color.                 \\ \hline
30 & \emph{Stock}           &   \cite{stock}     & To show the pattern of 5 stock prices over roughly 10 years from 2000 to 2010.  The objective is to show how each stock performed individually and comparatively with others. \\ \hline
31 & \emph{Airline}         &    \cite{airlines}  & To show and compare the changes in revenue of 4 airline companies over one year and demonstrate when sales for all went down (during COVID).                                             \\ \hline
32 & \emph{IMDB}            &    \cite{imdb}    & To show and compare the pattern in revenues made in the US vs. worldwide of a set of movies with varying IMDB ratings. Both graphs show positive correlations.         \\ \hline
33 & \emph{Visitors}        &   \cite{visitors}     & To show and compare patterns in the number of visitors received by 4 museums on a set of days. Show outlier cases.     \\ \hline
34 & \emph{Spotify}         &    \cite{spotify}    & To show and compare the streaming pattern of the 3 popular songs on Spotify over April 2017.                                          \\ \hline
35 & \emph{Stock}           &    \cite{stock}    & To show and compare the prices of 4 different stocks over roughly a decade by the juxtaposition between 2000-2010.   \\ \hline
36 & \emph{Activity-covid}  &    \cite{mccarthy2021physical}    & To show and compare the change in time people from 4 age groups spent on different activities before and after the Covid-19 pandemic. The graphs are shown in three categories: grooming, exercise, and mobile.     \\ \hline
37 & \emph{Unemployed}      &  \cite{unemployment}     & To show and compare patterns in the yearly count of people unemployed in specific industries over roughly 3 decades in three different categories.                                    \\ \hline
38 & \emph{Covid-3}         &    \cite{ritchie2020coronavirus}    & To show and compare patterns in the monthly count of (Covid-19) cases, deaths, and hospitalizations in 3 counties over one year.    \\ \hline
39 & \emph{Income}          &   \cite{kohavi1996scaling}     & To show correlation and compare two types of employees’ average weekly incomes and their growth rate over 16 years. There is a positive correlation between dollars per week and years, but the same distribution is not persistent in a change of income from the previous year.                                 \\ \hline
40 & \emph{Working}         &    \cite{maier2016descriptive}    & To show the pattern and compare the yearly percentage of the population in the workforce of six countries over roughly 3 decades, demonstrated by two graphs: Europe and North America.         \\ \hline
41 & \emph{12-cat products} &    \cite{12product}    & To show produces output(max-min) belonging to 12 categories using color choices that resemble the colors of their corresponding produce categories.                   \\ \hline
42 & \emph{Afghan}          &    \cite{Afghanistan}    & To show the census results and trend in Afghanistan population over roughly 7 decades with a design that makes it easy to tell a trend.                                           \\ \hline
43 & \emph{TempChange}      &    \cite{tempchange}    & To show the trend in the changes of US temperature over roughly one and a half-century using 2 colors indicating an increase or decrease in temperature to enable users to easily tell a trend.                                                     \\ \hline
44 & \emph{Budget}          &    \cite{movie}    & To show and compare the budget of different genres (min-max) of films by indicating films with very high budgets.                          \\ \hline
45 & \emph{Electricity}     &   \cite{electricity}     & To show the pattern and compare the prices of monthly electricity bills paid by a   household over a year and identify the period where the bill is higher or lower.                                \\ \hline
46 & \emph{Movie}           &    \cite{movie}    & To show and compare the average revenues made in the US and worldwide by movies of various types with texts that increase readability.                                           \\ \hline
47 & \emph{ProgProficiency} &   \cite{programming}     & To show pattern and compare 4 groups of people’s proficiency in various programming languages. The proficiency increases with their education and experience.                        \\ \hline
48 & \emph{Textured-Alt}    &    \cite{textures}    & To show and compare (min-max) the count of likes and dislikes received by 4 YouTube videos, categories each belonging to a distinct channel.                                              \\ \hline
49 & \emph{Textured}        &    \cite{textures}    & To show and compare (min-max)the count of likes and dislikes received by 4 YouTube videos, each belonging to a different channel.      \\ \hline
50 & \emph{Visitors}        &   \cite{visitors}     & To show the pattern and compare (min-max) the average monthly count of visitors received by 4 museums in a given year.                       \\ \hline
51 & \emph{Revenue}         &   \cite{movie}     & To show patterns and compare the budget and the gross revenue of movies of various types/genres.                                          \\ \hline
52 & \emph{Success}         &    \cite{success}    & To show and compare various countries’ opinions on how much a specific factor plays a part in contributing to an individual’s success.                                            \\ \hline
53 & \emph{Avg sales}       &  \cite{avgsales}  & To show patterns and compare the sales figures or average monthly sales and distribution and patterns of 4 locations over a year. \\ \hline
54 & \emph{Media}           &   \cite{dixon2022average}   & To show patterns and compare the time a user spends using social media apps and entertainment apps in a given week.             \\ \hline
55 & \emph{Snapchat\&Ins}         &  \cite{dixon2022average}   & To show the screen time pattern a user spends on Snapchat and Instagram in a given week and compare patterns between them side by side.  \\ \hline
56 & \emph{Products}        &    \cite{owidagriculturalproduction}    & To show min-max and compare the output of various foods belonging to 3 categories using colors that resemble their corresponding food’s color in real life.                               \\ \hline
57 & \emph{Farms}           &   \cite{farms}     & To show and compare (min-max) the output harvest (produce) of 4 types of fruits in 6 locations.                              \\ \hline
58 & \emph{ProgPopularity}  &   \cite{programming}     & To show the pattern and compare the monthly market share of 6 programming languages in 2021, demonstrated by quarters.   \\ \hline
59 & \emph{Q\&A} &    \cite{akande2020dataset}    & To show the pattern and compare the percentage of respondents showing various   attitudes towards a set of problems                  \\ \hline
60 & \emph{Stock}           &    \cite{stock}    & To show and compare the prices of 5 stocks in the January of 2005-2008.  Google stock has the highest price.  
\\ \hline
\end{longtable}
\twocolumn

\end{document}